\documentclass[twocolumn,showpacs,preprintnumbers,amsmath,amssymb]{revtex4}
\usepackage{graphicx}
\usepackage{dcolumn}
\usepackage{bm}
\usepackage{hyperref}
\usepackage{xcolor}

\newcommand{\bef}{\begin{figure}}
\newcommand{\eef}{\end{figure}}

\newcommand{\be}{\begin{equation}}
\newcommand{\ee}{\end{equation}}
\newcommand{\bea}{\begin{eqnarray}}
\newcommand{\eea}{\end{eqnarray}}
\usepackage{natbib}

\begin{document}

\title{Probing $\alpha$ clustering in $^{12}\mathrm{C}$ at energies available at the Cooling Storage Ring facility using the Jet AA Microscopic Transport Model}

\author{Subhash Singha}
\email {subhash@impcas.ac.cn}
\affiliation{Institute of Modern Physics Chinese Academy of Sciences, Lanzhou 730000, China\\
State Key Laboratory of Heavy Ion Science and Technology,\\
Institute of Modern Physics Chinese Academy of Sciences, Lanzhou 730000, China
}

\date{\today}

\begin{abstract}

 I investigate the sensitivity of low-energy nuclear collisions to intrinsic nuclear structure by studying the interplay between initial-state geometry and final-state observables in C+C and C+Pb collisions at $\sqrt{s_{\mathrm{NN}}}=2.36$~GeV, relevant for experiments at the Cooling Storage Ring (CSR) facility in Lanzhou and forthcoming experiments at the High Intensity heavy-ion Accelerator Facility (HIAF) in Huizhou. Calculations are performed within the Jet AA Microscopic Transport Model (JAM) using Woods--Saxon and triangular $\alpha$-clustered configurations for the $^{12}\mathrm{C}$ nucleus. The initial geometry is characterized in terms of transverse size, compactness, eccentricities, and their ensemble-averaged fluctuations. I find that $\alpha$ clustering leads to a more compact participant configuration than the Woods--Saxon case, while transverse-size and eccentricity fluctuations show only weak sensitivity to clustering. At this beam energy, radial observables remain sensitive to geometric compactness, with the ensemble-averaged proton mean transverse momentum $\langle p_T \rangle$ enhanced for the $\alpha$-clustered configuration, whereas pion observables show little sensitivity. The anisotropic response is examined using flow harmonic coefficients. While the flow harmonics and their event-by-event fluctuations exhibit no appreciable differences between the Woods--Saxon and $\alpha$-clustered configurations, normalized symmetric cumulants of the initial-state eccentricities show sensitivity to clustering, whereas the corresponding symmetric cumulants among the final-state flow harmonics do not exhibit a comparably strong separation. These results indicate that the dynamical evolution at $\sqrt{s_{\mathrm{NN}}}=2.36$~GeV substantially weakens the imprint of the initial clustered geometry on anisotropic-flow observables, while radial observables remain promising probes of $\alpha$ clustering in low-energy nuclear collisions.

\end{abstract}

\maketitle

\section{Introduction}

In recent years, a new direction in relativistic heavy-ion physics has emerged that exploits the geometric and structural properties of the colliding nuclei as tools to probe both nuclear structure and the collective response of the produced medium. Experimental measurements have demonstrated that intrinsic nuclear deformation can leave observable imprints on final-state momentum anisotropies. In particular, the STAR Collaboration has reported clear signatures of the quadrupole deformation of uranium nuclei in U+U collisions at RHIC energies~\cite{STAR:2024wgy,STAR:2025elk}, while measurements in Xe+Xe collisions at the LHC~\cite{ALICE:2018lao,CMS:2019cyz,ATLAS:2019dct} have revealed sensitivity to the prolate deformation of the xenon nucleus through deviations from expectations based on spherical nuclei.

These developments have motivated growing interest in using light nuclei with pronounced intrinsic structure as controlled probes of initial-state geometry in nuclear collisions. In this context, collisions involving nuclei such as $^{12}\mathrm{C}$ and $^{16}\mathrm{O}$ are particularly intriguing, as they are predicted to exhibit strong $\alpha$-cluster correlations in their ground-state wave functions~\cite{Broniowski:2013dia,He:2014iqa,Rybczynski:2017nrx,Zhang:2017xda,Guo:2017tco,Yao:2025mqh,Ma:2020vqe,Zhang:2024vkh,Fan:2025sfd,Hu:2025eid,Prasad:2024ahm,MenonKavumpadikkalRadhakrishnan:2025apq,MenonKavumpadikkalRadhakrishnan:2024twu,YuanyuanWang:2024sgp,Zhao:2024feh,Shafi:2025feq,Li:2025hae}. The intrinsic structure of $^{12}\mathrm{C}$, and in particular the nature of its Hoyle state, has been the subject of extensive theoretical and experimental investigations. While \emph{ab initio} calculations based on effective field theory have suggested that the Hoyle state may resemble a linear chain of three $\alpha$ clusters~\cite{Epelbaum:2011md}, more recent experimental measurements favor a predominantly triangular $\alpha$-cluster configuration~\cite{Marin-Lambarri:2014zxa,Shen:2022bak}, indicating that its underlying geometry remains an active subject of study.

Recent experimental results from O+O collisions at RHIC~\cite{STAR:2025ivi} and the LHC~\cite{ALICE:2024nqd,ALICE:2024apz,ATLAS:2025nnt,ATLAS:2025ztg,CMS:2025tga}, together with complementary theoretical studies, suggest that small-system collisions may retain sensitivity to such subnucleonic or cluster-scale geometric features. These observations open the possibility of accessing clustering effects through final-state observables in relativistic nuclear collisions. Since $\alpha$ clustering is an intrinsic feature of the initial coordinate-space geometry of the nucleus, its experimental manifestation necessarily relies on the efficiency with which these spatial correlations are dynamically translated into measurable momentum-space signatures.

At ultra-relativistic collision energies, the connection between nuclear geometry and final-state observables is commonly interpreted within a hydrodynamic framework. In this picture, initial-state spatial anisotropies and density inhomogeneities are efficiently converted into momentum-space correlations through the collective expansion of a nearly equilibrated medium. The system’s response is predominantly governed by pressure gradients and transport properties, enabling nuclear structure effects to be investigated in a controlled manner via their influence on initial eccentricities and density profiles.

At lower collision energies, however, the dynamical evolution of the system is qualitatively different. Baryon stopping, finite net-baryon density, mean-field potentials, and hadronic rescattering play a central role and can substantially modify the mapping between the initial geometry and final-state observables~\cite{Videbaek:2005qn,Mohs:2019xvz,Ivanov:2010cu,Isse:2005nk,Nara:2021fuu,Chen:2017cjw}. Consequently, it is not a priori evident whether, or to what extent, nuclear structure effects such as $\alpha$ clustering survive the subsequent evolution and remain observable. Determining how initial geometric features are preserved in this regime therefore constitutes an important open question and is essential for assessing the sensitivity of low-energy heavy-ion collisions to nuclear structure.

Recent and forthcoming experimental programs provide new opportunities to explore these questions. Experiments involving C+C and C+Pb collisions at $\sqrt{s_{\mathrm{NN}}}=2.36$~GeV have recently commenced at the Cooling Storage Ring (CSR) facility in Lanzhou~\cite{Lu:2016htm,Xia:2002xpu,Mao:2020rlb,Zhu:2021soc,Liu:2023xhc,Sahoo:2023fzz,Hu:2023niz,Wu:2025dev,Hong2026}. In addition, the upcoming High Intensity heavy-ion Accelerator Facility (HIAF)~\cite{Yang:2013yeb,Zhou:2022pxl,Yang:2025sni,HIAF:web} in Huizhou will enable similar studies with a variety of collision systems, including O+O and U+U. Significant efforts are also underway at the LHC, where collisions of nuclei with non-spherical or exotic configurations, such as deformed Ne+Ne systems~\cite{Giacalone:2024luz,ATLAS:2025nnt}, have been performed recently. Together, these programs open a new avenue for systematically investigating nuclear structure effects across a wide range of collision energies.

\begin{figure}
\begin{center}
\includegraphics[scale=0.35]{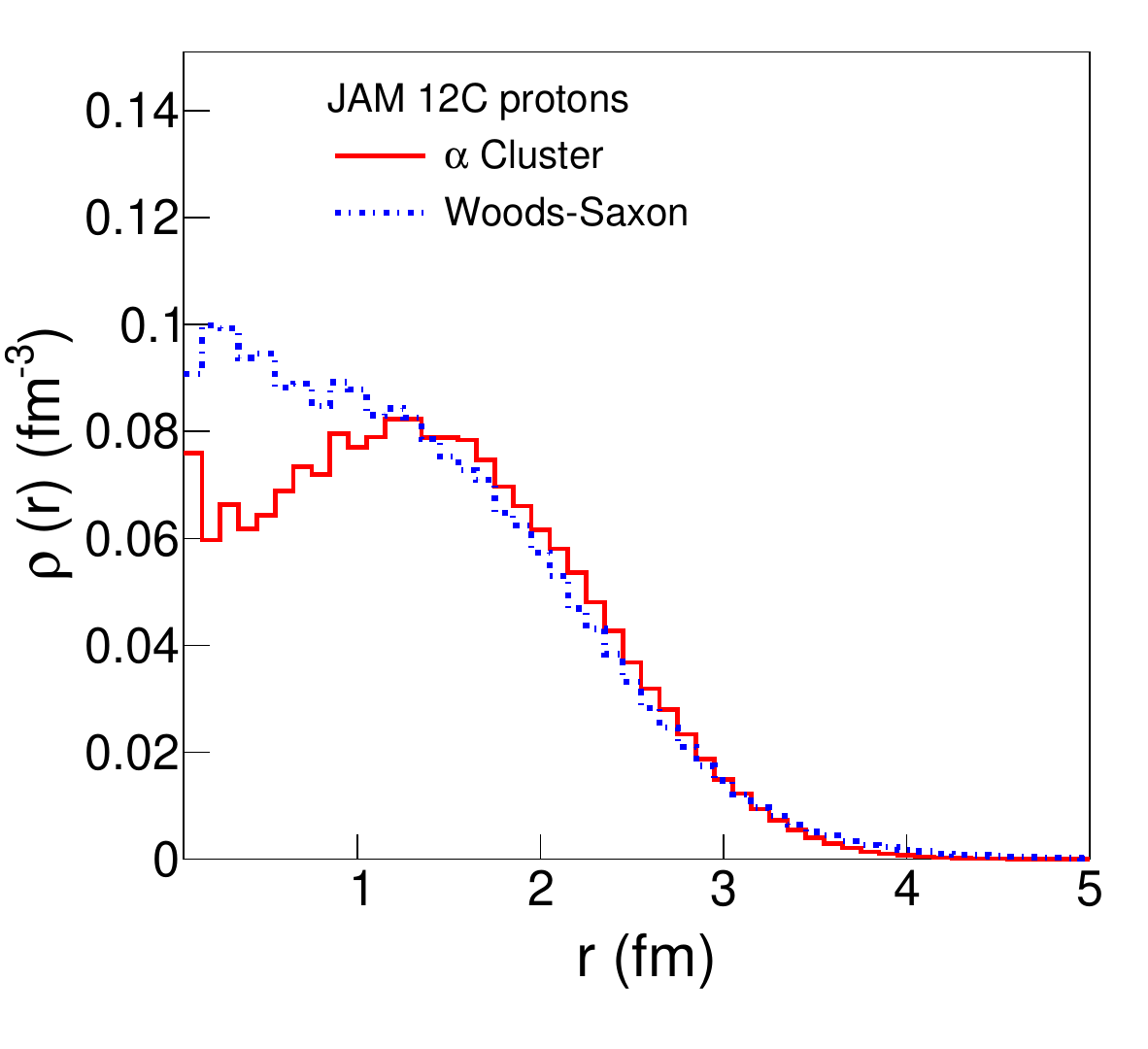}
\caption{Radial distribution of the point-proton density of $^{12}\mathrm{C}$ obtained from JAM initial configurations for Woods--Saxon and $\alpha$-clustered nuclei. The densities are normalized to the total proton number.}
\label{fig_charge_radius}
\end{center}
\end{figure} 

\section{JAM Framework}
In this work, I employ the Jet AA Microscopic Transport Model (JAM), a hadronic transport framework developed for the simulation of relativistic nuclear collisions over a broad range of beam energies~\cite{Nara:1999dz,Hirano:2012kj,Isse:2005nk,Nara:2019qfd,Nara:2020ztb,Nara:2021fuu}. In JAM, the initial positions of nucleons inside each nucleus are sampled according to a prescribed nuclear density distribution, after which the collision dynamics are modeled as a sequence of binary hadron--hadron interactions combined with mean-field effects. The model describes the full time evolution of the system, from the initial nuclear overlap through the dense interaction stage to the final hadronic freeze-out.

For the present study, JAM (version 2.58)~\cite{JAM:github}is used in its mean-field mode, which incorporates nuclear equation-of-state effects through a momentum-dependent potential that modifies the propagation of hadrons. This approach is essential at low collision energies, where baryon stopping and mean-field interactions play a significant role in shaping collective dynamics. Events are generated for C+C and C+Pb collisions at $\sqrt{s_{\mathrm{NN}}}=2.36$~GeV, both with and without $\alpha$ clustering in the $^{12}\mathrm{C}$ nucleus, where the implementation of clustering will be discussed in detail in the following section. I use the MD2 equation of state~\cite{Nara:2020ztb}, characterized by a nuclear incompressibility of $K=380$~MeV, which has been shown to successfully reproduce collective flow observables in low-energy heavy-ion collisions at RHIC and lower energies. The time evolution is followed up to $100~\mathrm{fm}/c$ with a time step of $1~\mathrm{fm}/c$.

\subsection{Clustering implementation}

To investigate the role of nuclear structure effects, two distinct initial configurations are employed for the $^{12}\mathrm{C}$ nucleus. In the reference configuration, nucleon positions are sampled from a spherical Woods--Saxon distribution,
\begin{equation}
\rho(r)=\frac{\rho_0}{1+\exp[(r-R)/a]},
\end{equation}
using the default parameterization implemented in JAM2 and inherited from Pythia8, with
\begin{equation}
R = 1.12A^{1/3} - 0.86A^{-1/3}~\mathrm{fm}, \qquad
a = 0.54~\mathrm{fm}.
\end{equation}
This corresponds to $R = 2.19$ fm for $^{12}$C and $R = 6.49$ fm for $^{208}$Pb.

In the alternative configuration, $^{12}\mathrm{C}$ is modeled as an $\alpha$-clustered system consisting of three $\alpha$ clusters arranged at the vertices of an equilateral triangle in the transverse plane. This triangular configuration is motivated by nuclear structure studies, which have long predicted a pronounced three-$\alpha$ cluster structure in the ground state of $^{12}\mathrm{C}$~\cite{hafstad1938alpha,Harrington:1966zz,ikeda1968systematic,Noble:1970nxx,Bertsch:1971abe}. Both transport and hydrodynamic models have been extensively employed to explore possible experimental signatures of nuclear clustering in heavy-ion collisions over a wide range of beam energies~\cite{Broniowski:2013dia,He:2014iqa,Rybczynski:2017nrx,Zhang:2017xda,Guo:2017tco,Ma:2020vqe,Liu:2023gun,Yao:2025mqh}.

The centers of the $\alpha$ clusters are separated by a fixed distance $d_{\alpha\alpha}$, which sets the characteristic length scale of the clustered geometry. Within each $\alpha$ cluster, the four constituent nucleons are distributed around the corresponding cluster center according to a Gaussian profile, $\rho_{\alpha}(\mathbf{r}) \propto \exp\!\left(-\mathbf{r}^{2}/2\sigma^{2}\right)$,
where the width parameter $\sigma$ controls the spatial localization of nucleons inside the cluster. In the present implementation, the values of $d_{\alpha\alpha}$ and $\sigma$ are fixed for all events, while the overall orientation of the triangular configuration is randomized on an event-by-event basis. To prevent unphysical overlap between nucleons, a hard-core distance ($0.9$ fm) is imposed between all nucleon pairs.

The parameters of the $\alpha$-clustered configuration ($d_{\alpha\alpha}$=3.1 fm and $\sigma$=0.8) are chosen such that the resulting point-proton density distribution and root-mean-square radius of $^{12}\mathrm{C}$ remain comparable to those obtained from the Woods--Saxon configuration. This ensures that the two initial conditions differ primarily in their internal spatial organization and short-range correlations, rather than in their global size or average density, thereby enabling a controlled study of clustering-induced geometric effects. With this setup, JAM provides a consistent framework in which the impact of $\alpha$ clustering on the initial geometry and its subsequent influence on final-state observables can be systematically investigated while keeping the underlying transport dynamics and equation of state fixed.

The proton density distribution of $^{12}\mathrm{C}$ is constructed from the initial-state nucleon configurations generated in JAM by accumulating proton coordinates over all events to obtain a spherically averaged density,
\begin{equation}
\rho_p(r) = \frac{1}{4\pi r^2}\frac{dN_p}{dr},
\end{equation}
which is normalized according to
\begin{equation}
\int d^3r\, \rho_p(r) = 6.
\end{equation}
No additional smearing associated with the finite proton charge radius is applied, such that the resulting distribution corresponds to the point-proton density. The same procedure is used consistently for both Woods--Saxon and $\alpha$-clustered initial configurations.


Differences between the two configurations arise primarily from their internal spatial structure rather than from the global density scale. In particular, $\alpha$ clustering modifies short-range correlations and local proton packing while  the radial density remains approximately similar between the two configurations, with expected deviations at small $r$ due to clustering. This is reflected in the fact that the root-mean-square (RMS) radius of the Woods--Saxon configuration of $^{12}\mathrm{C}$ is $r_{\mathrm{rms}}\simeq 2.4$~fm, while that of the $\alpha$-clustered configuration is $r_{\mathrm{rms}}\simeq 2.3$~fm, corresponding to a difference of less than $4\%$. This demonstrates that the observed differences in initial-state compactness and in the final-state proton mean transverse momentum do not originate from trivial changes in the overall nuclear size or charge density, but instead reflect differences in the microscopic spatial organization of the nucleons.

\begin{figure*}
\includegraphics[scale=0.65]{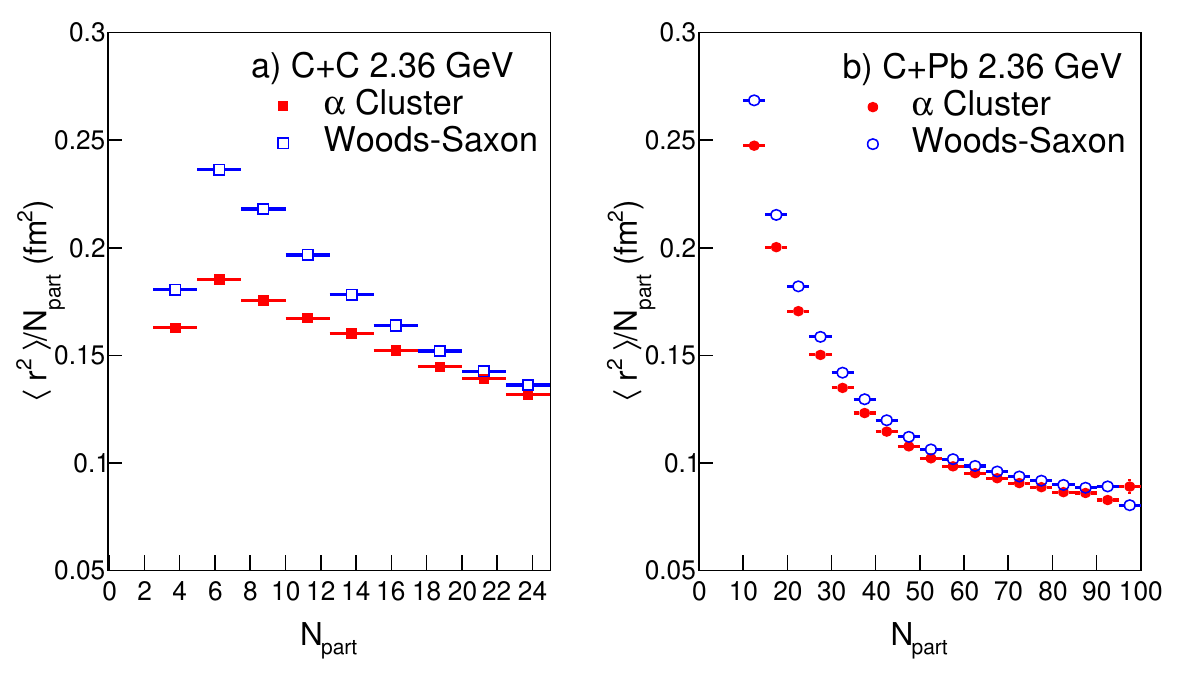}
\caption{The transverse size ($\langle r^{2} \rangle$) as a function of
$N_{\mathrm{part}}$ for Woods--Saxon and triangular $\alpha$-clustered initial
configurations for C+C (panel a) and C+Pb (panel b) collisions at $\sqrt{s_{\mathrm{NN}}}=2.36$~GeV.
}
\label{fig_trans_radius}
\end{figure*} 

\begin{figure*}
\includegraphics[scale=0.65]{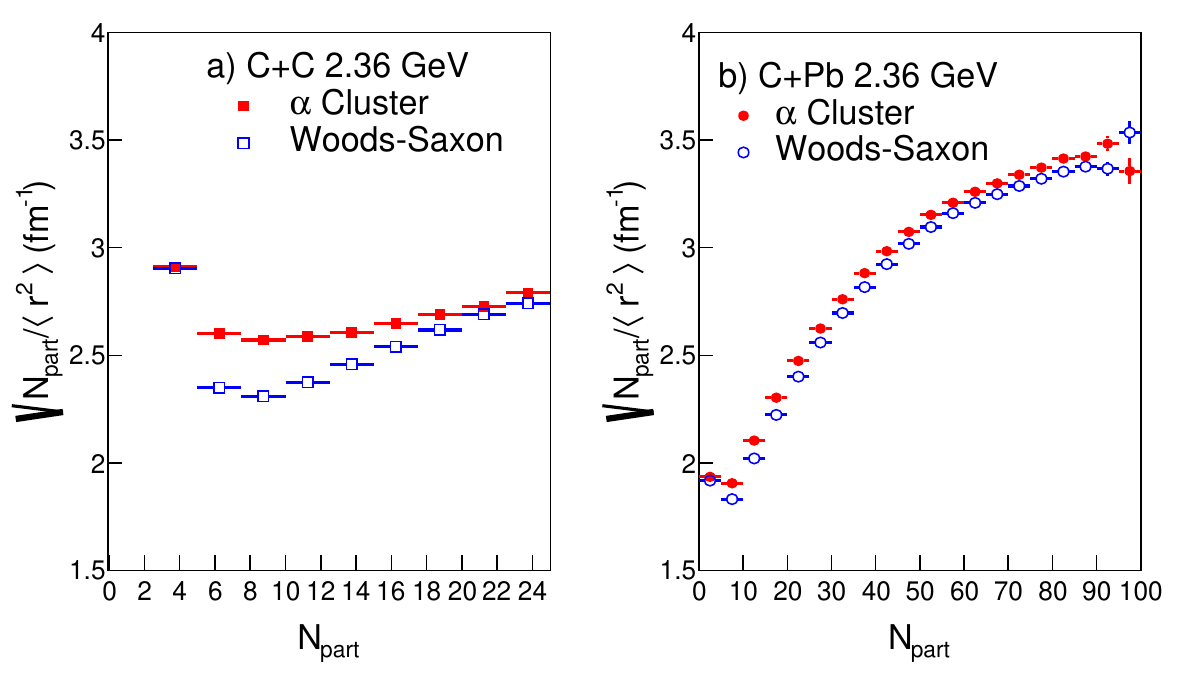}
\caption{The compactness ($C$) as a function of
$N_{\mathrm{part}}$ for Woods--Saxon and triangular $\alpha$-clustered initial
configurations for C+C (panel a) and C+Pb (panel b) collisions at $\sqrt{s_{\mathrm{NN}}}=2.36$~GeV.
}
\label{fig_trans_compactness}
\end{figure*} 

\begin{figure*}
\includegraphics[scale=0.65]{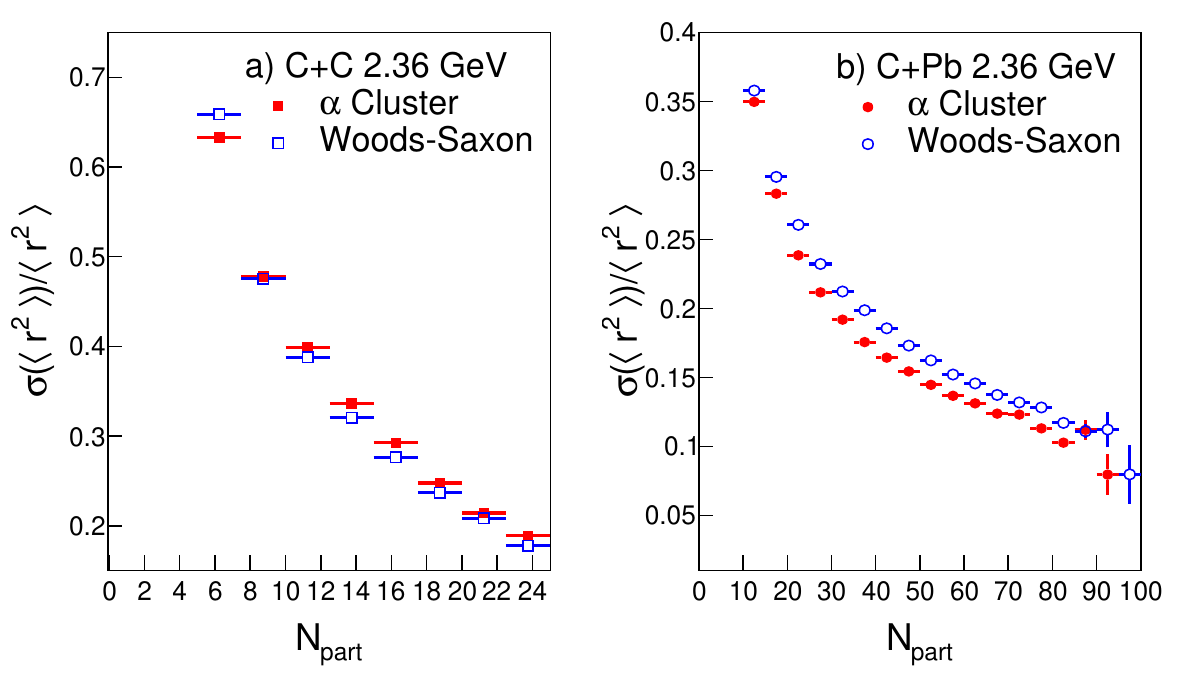}
\caption{The transverse-size fluctuation as a function of
$N_{\mathrm{part}}$ for Woods--Saxon and triangular $\alpha$-clustered initial
configurations for C+C (panel a) and C+Pb (panel b) collisions at $\sqrt{s_{\mathrm{NN}}}=2.36$~GeV.
}
\label{fig_trans_radius_fluc}
\end{figure*} 


\section{Results and discussions}

\subsection{Correlation in initial-state geometries}
\subsubsection{Transverse radius correlation}
To characterize the geometric properties of the initial participant configuration, I consider several complementary observables constructed from the transverse coordinates of participant nucleons~\cite{Broniowski:2007nz,Broniowski:2007nz}. The average transverse size is quantified by the second moment
\begin{equation}
\langle r^{2} \rangle
=
\frac{1}{N_{\mathrm{part}}}
\sum_{i=1}^{N_{\mathrm{part}}}
\left[(x_i-\bar{x})^{2}+(y_i-\bar{y})^{2}\right],
\end{equation}
where $(\bar{x},\bar{y})$ denotes the transverse center of mass of the participants in a given event. To remove the trivial scaling with the number of participants, I also study the scaled transverse size $\langle r^{2} \rangle / N_{\mathrm{part}}$. The corresponding compactness of the participant zone is defined as
\begin{equation}
C \equiv \sqrt{\frac{N_{\mathrm{part}}}{\langle r^{2} \rangle}},
\end{equation}
which provides a measure of the inverse transverse length scale governing the initial pressure gradients. Fluctuations of the transverse size are quantified by the relative width
\begin{equation}
\frac{\sigma(\langle r^{2} \rangle)}{\langle r^{2} \rangle},
\end{equation}
where the variance of $\langle r^{2} \rangle$ is defined as
\begin{equation}
\sigma(\langle r^{2} \rangle)
=
\sqrt{
\left\langle (\langle r^{2} \rangle)^{2} \right\rangle
-
\left\langle \langle r^{2} \rangle \right\rangle^{2}
},
\end{equation}
with $\langle \cdots \rangle$ denoting the ensemble average over events at fixed $N_{\mathrm{part}}$.

The left and right panels in Figures~\ref{fig_trans_radius},~\ref{fig_trans_compactness} and ~\ref{fig_trans_radius_fluc} show the scaled transverse size ($\langle r^{2} \rangle / N_{\mathrm{part}}$), compactness ($C$), and the corresponding transverse-size fluctuation strength as functions of $N_{\mathrm{part}}$ for Woods--Saxon and triangular $\alpha$-clustered initial configurations, respectively, in both C+C and C+Pb collisions at $\sqrt{s_{\mathrm{NN}}}=2.36$~GeV.

In C+C collisions, the $\alpha$-clustered configuration exhibits a systematically smaller scaled transverse size than the Woods--Saxon case, indicating a more compact spatial arrangement of the participant nucleons. This reduction in $\langle r^{2} \rangle / N_{\mathrm{part}}$ is accompanied by a corresponding enhancement of the compactness over the full $N_{\mathrm{part}}$ range. In contrast, the relative transverse-size fluctuations show a weaker sensitivity to clustering, with comparable magnitudes for Woods--Saxon and $\alpha$-clustered configurations, while the magnitude of the fluctuations decreases with increasing $N_{\mathrm{part}}$. The observed pattern demonstrates that, in symmetric light--light collisions, $\alpha$ clustering primarily shifts the average transverse scale of the system rather than significantly enhancing transverse-size fluctuations.

A different behavior is observed in C+Pb collisions. While the average compactness remains larger for the $\alpha$-clustered configuration, reflecting the increased local packing of nucleons in the carbon projectile, the relative transverse-size fluctuations are comparable or slightly larger for the Woods--Saxon case. In this asymmetric system, fluctuations of the transverse size are dominated by the geometry and participant fluctuations on the Pb side, which reduces the relative impact of clustering in the lighter nucleus.
Consequently, $\alpha$ clustering continues to modify the mean transverse geometry but does not lead to an enhancement of transverse-size fluctuations. Overall, the comparison demonstrates that $\alpha$ clustering predominantly affects the average compactness of the participant zone, while its influence on transverse-size fluctuations is comparatively modest and most pronounced at low $N_{\mathrm{part}}$.

\begin{figure*}
\includegraphics[scale=0.65]{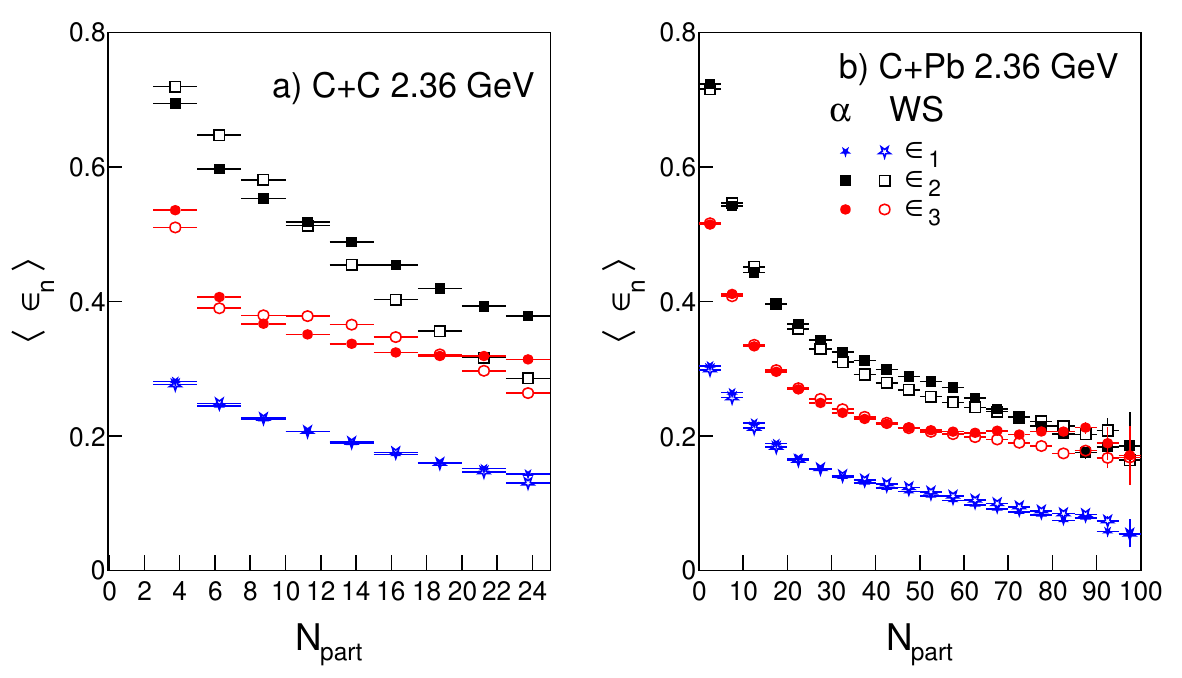}
\caption{Ensemble-averaged participant eccentricities ($\varepsilon_n$) as a function of $N_{\mathrm{part}}$ for Woods--Saxon and
triangular $\alpha$-clustered initial configurations for C+C (panel a) and C+Pb (panel b) collisions at
$\sqrt{s_{\mathrm{NN}}}=2.36$~GeV.
}
\label{fig_eccen_mean}
\end{figure*} 

\begin{figure*}
\includegraphics[scale=0.65]{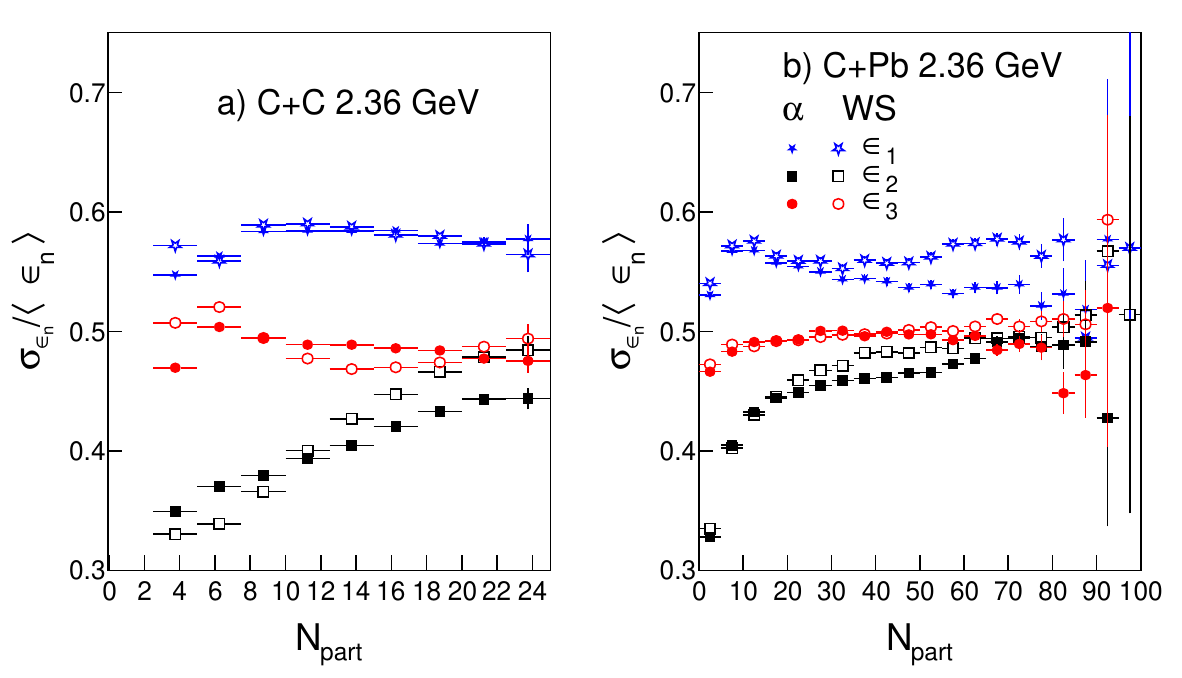}
\caption{Fluctuation of ensemble-averaged participant eccentricities ($\sigma(\varepsilon_n)/\langle\varepsilon_n \rangle$) as a function of $N_{\mathrm{part}}$ for Woods--Saxon and triangular $\alpha$-clustered initial configurations for C+C (panel a) and C+Pb (panel b) collisions at
$\sqrt{s_{\mathrm{NN}}}=2.36$~GeV.
}
\label{fig_eccen_fluc}
\end{figure*} 

\subsubsection{Eccentricity correlation}

In addition to transverse size observables, the anisotropic geometry of the initial participant configuration is usually characterized using event-by-event eccentricities $\varepsilon_n$. For each event, participant nucleons are identified and their transverse coordinates $(x_i,y_i)$ are evaluated relative to the participant center of mass. The eccentricity coefficients are defined using weighted moments of the participant distribution~\cite{Alver:2010gr}. 

For the dipole eccentricity ($n=1$), I use~\cite{Teaney:2010vd}
\begin{equation}
\varepsilon_1
=
\frac{\sqrt{
\left\langle r^3 \cos(\phi) \right\rangle^2
+
\left\langle r^3 \sin(\phi) \right\rangle^2
}}
{\left\langle r^3 \right\rangle}.
\end{equation}

For higher-order eccentricities ($n \geq 2$), I use~\cite{Alver:2010gr,Bhalerao:2011bp}
\begin{equation}
\varepsilon_n
=
\frac{\sqrt{
\left\langle r^2 \cos(n\phi) \right\rangle^2
+
\left\langle r^2 \sin(n\phi) \right\rangle^2
}}
{\left\langle r^2 \right\rangle},
\end{equation}
where $\phi=\tan^{-1}(y/x)$ is the azimuthal angle of a participant. These definitions follow standard prescriptions in studies of initial-state geometry and ensure sensitivity to long-wavelength shape deformations.

Event-averaged eccentricities are obtained as functions of $N_{\mathrm{part}}$ by averaging $\varepsilon_n$ over events within a given $N_{\mathrm{part}}$ bin. The fluctuation strength is quantified using the variance
\begin{equation}
\sigma(\varepsilon_n)
=
\sqrt{
\langle \varepsilon_n^2 \rangle
-
\langle \varepsilon_n \rangle^2
},
\end{equation}
as well as the corresponding relative fluctuation $\sigma(\varepsilon_n)/\langle \varepsilon_n \rangle$. These quantities characterize the intrinsic geometric variability of the initial state at fixed system size.

Figures~\ref{fig_eccen_mean} and~\ref{fig_eccen_fluc} present the average eccentricities $\varepsilon_n$ and their corresponding fluctuation strengths as functions of $N_{\mathrm{part}}$ for C+C (panel a) and C+Pb (panel b) collisions, respectively. For both the Woods--Saxon and $\alpha$-clustered configurations, the mean eccentricities follow the ordering $\varepsilon_2 > \varepsilon_3 > \varepsilon_1$, with comparable magnitudes in the two cases. The relative fluctuations of each $\varepsilon_n$ remain approximately constant as functions of $N_{\mathrm{part}}$ for $\varepsilon_1$ and $\varepsilon_3$, while the fluctuation of $\varepsilon_2$ increases with increasing $N_{\mathrm{part}}$. For both C+C and C+Pb collisions, the $\varepsilon_n$ fluctuations follow the ordering $\sigma(\varepsilon_1) > \sigma(\varepsilon_3) > \sigma(\varepsilon_2)$.

\begin{figure*}
\includegraphics[scale=0.65]{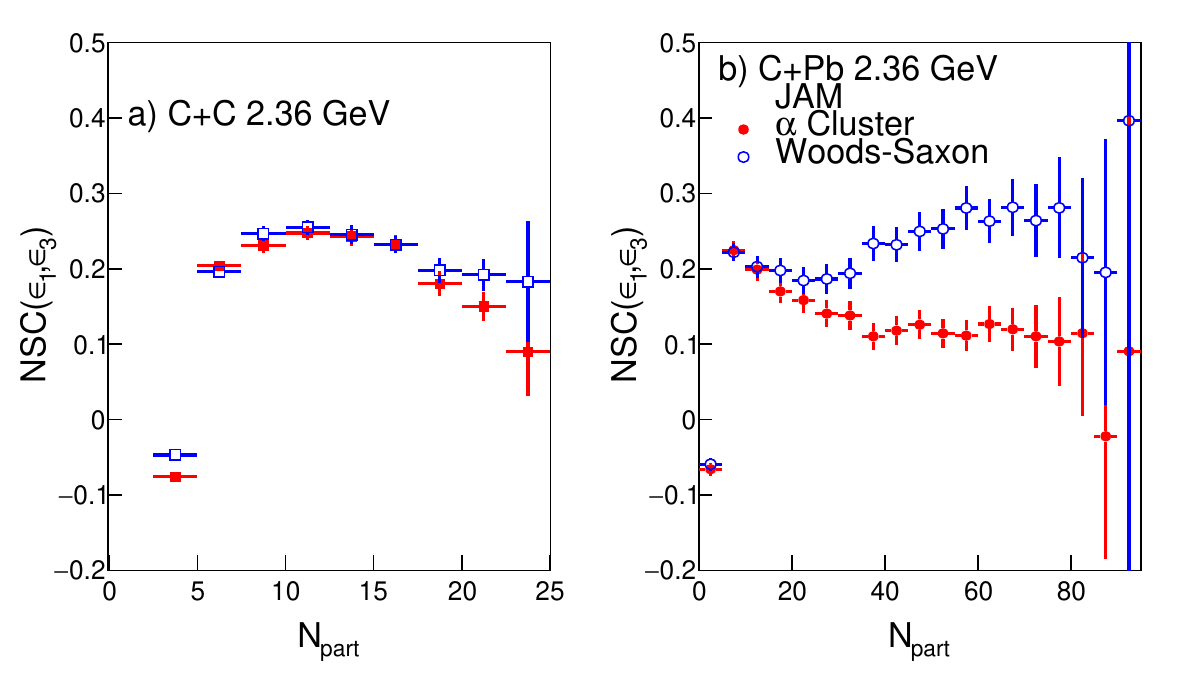}
\caption{Symmetric cumulants of eccentricities $\mathrm{NSC}\{\varepsilon_1,\varepsilon_3\}$ a function of $N_{\mathrm{part}}$ for Woods--Saxon and triangular $\alpha$-clustered initial configurations for C+C (panel a) and C+Pb (panel b) collisions at $\sqrt{s_{\mathrm{NN}}}=2.36$~GeV. 
}
\label{fig_eccen_nsc13}
\end{figure*} 


\begin{figure*}
\includegraphics[scale=0.65]{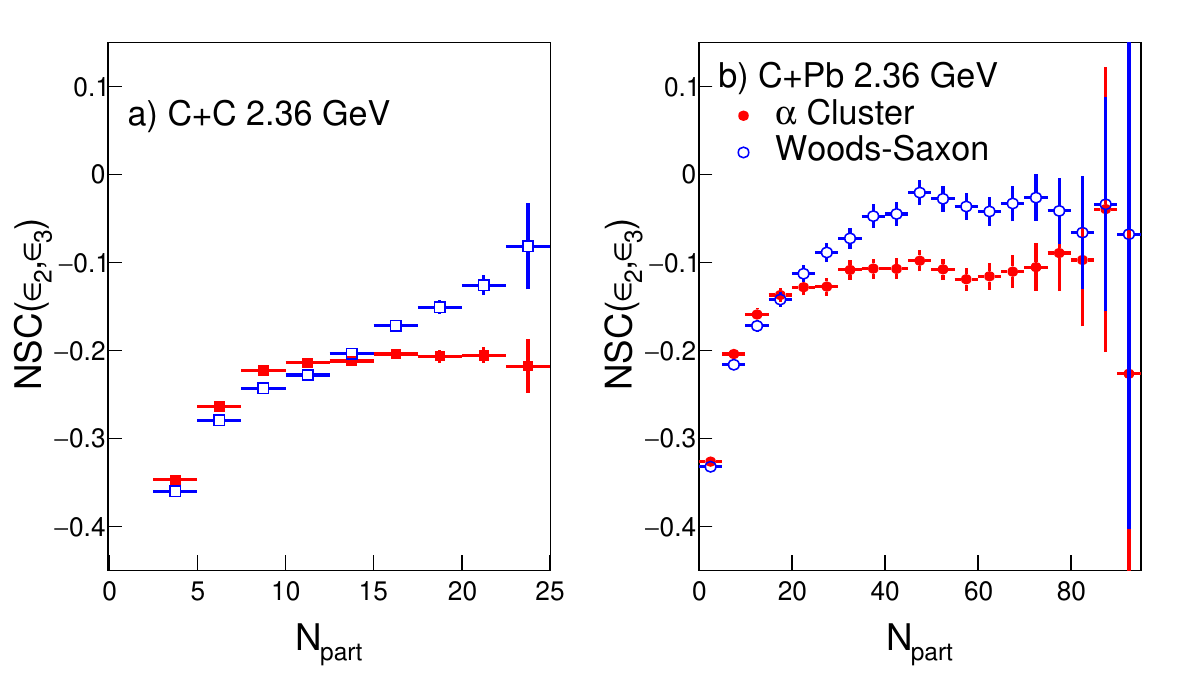}
\caption{Normalized symmetric cumulants of eccentricities $\mathrm{NSC}\{\varepsilon_2,\varepsilon_3\}$ a function of $N_{\mathrm{part}}$ for Woods--Saxon and triangular $\alpha$-clustered initial configurations for C+C (panel a) and C+Pb (panel b) collisions at $\sqrt{s_{\mathrm{NN}}}=2.36$~GeV. 
}
\label{fig_eccen_nsc23}
\end{figure*} 

Correlations between different harmonic components of the initial geometry are investigated using mixed moments, specifically $\langle \varepsilon_1^{2}\varepsilon_3^{2} \rangle$ and $\langle \varepsilon_2^{2}\varepsilon_3^{2} \rangle$, together with the corresponding single-harmonic moments $\langle \varepsilon_1^{2} \rangle$, $\langle \varepsilon_2^{2} \rangle$, and $\langle \varepsilon_3^{2} \rangle$. These moments quantify correlations between different geometric deformations beyond their individual magnitudes. For two harmonics $m$ and $n$, the symmetric cumulant (SC) is defined as~\cite{Bilandzic:2013kga,DiFrancesco:2016srj}
\begin{equation}
\mathrm{SC}\{\varepsilon_m,\varepsilon_n\}
=
\langle \varepsilon_m^{2}\varepsilon_n^{2} \rangle
-
\langle \varepsilon_m^{2} \rangle
\langle \varepsilon_n^{2} \rangle ,
\end{equation}
and the corresponding normalized symmetric cumulant (NSC) is defined as
\begin{equation}
\mathrm{NSC}\{\varepsilon_m,\varepsilon_n\}
=
\frac{
\mathrm{SC}\{\varepsilon_m,\varepsilon_n\}
}{
\langle \varepsilon_m^{2} \rangle
\langle \varepsilon_n^{2} \rangle
}.
\end{equation}
Here, the angular brackets denote ensemble averages over events at fixed $N_{\mathrm{part}}$. Positive (negative) values of the SC and NSC indicate correlated (anticorrelated) fluctuations of $\varepsilon_m^{2}$ and $\varepsilon_n^{2}$. Unlike the unnormalized SC, the NSC reduces the dependence on the overall magnitudes of the eccentricities and therefore provides a more direct measure of the relative correlation strength.

\begin{figure*}
\includegraphics[scale=0.5]{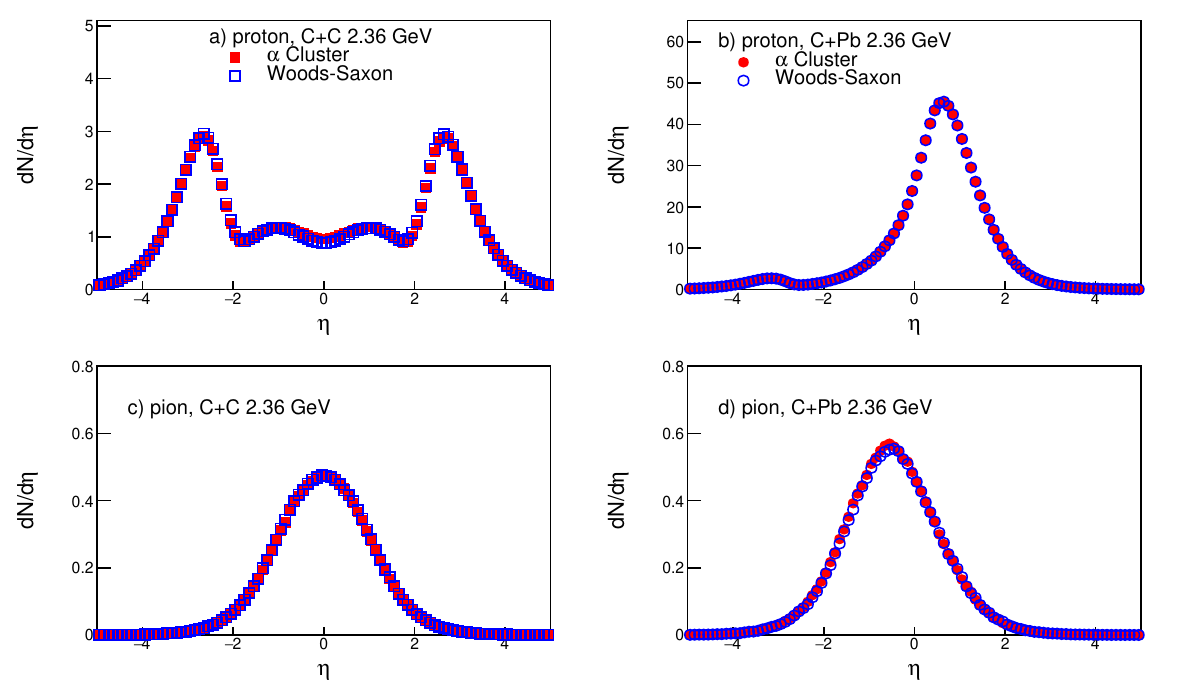}
\caption{$dN/d\eta$ of protons and pions for Woods--Saxon and triangular $\alpha$-clustered initial configurations for C+C and C+Pb collisions at $\sqrt{s_{\mathrm{NN}}}=2.36$~GeV, integrated over all $N_{\mathrm{part}}$.
}
\label{fig_dndeta}
\end{figure*} 

\begin{figure*}
\includegraphics[scale=0.5]{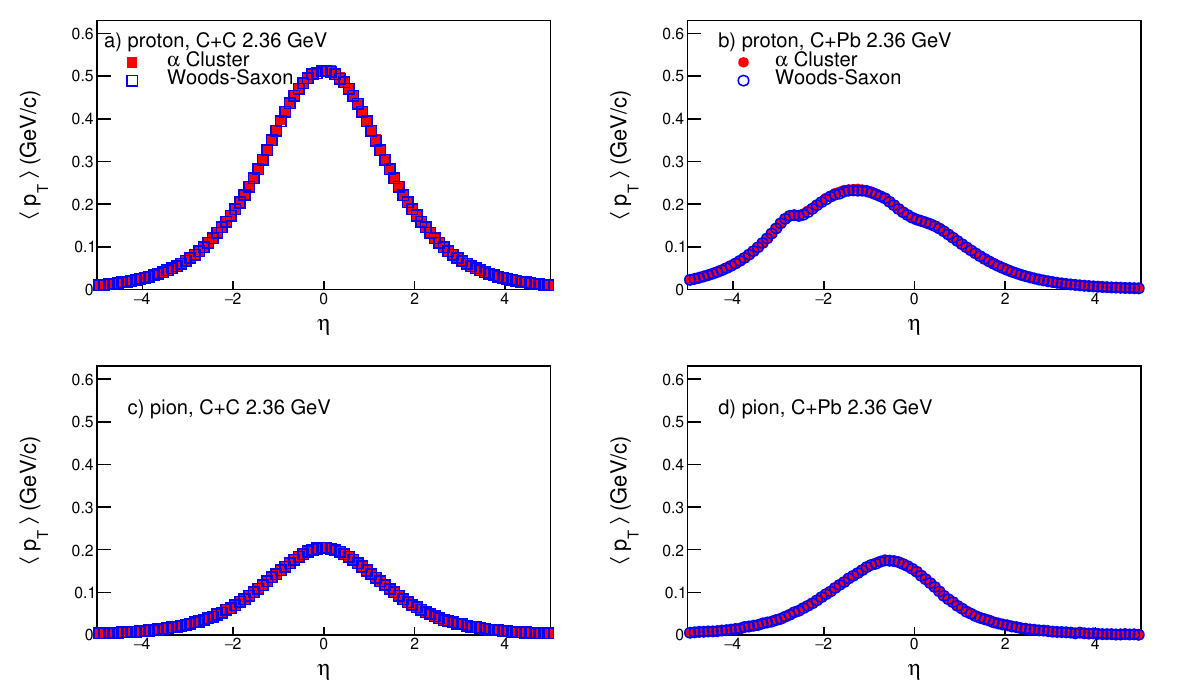}
\caption{$\langle p_T \rangle$ versus $\eta$ of protons and pions for Woods--Saxon and triangular $\alpha$-clustered initial configurations for C+C and C+Pb collisions at $\sqrt{s_{\mathrm{NN}}}=2.36$~GeV, integrated over all $N_{\mathrm{part}}$.
}
\label{fig_meanptEta}
\end{figure*} 

The normalized symmetric cumulants, $\mathrm{NSC}\{\varepsilon_1,\varepsilon_3\}$ and $\mathrm{NSC}\{\varepsilon_2,\varepsilon_3\}$, are presented in Figs.~\ref{fig_eccen_nsc13} and~\ref{fig_eccen_nsc23}, respectively. For $\mathrm{NSC}\{\varepsilon_1,\varepsilon_3\}$, a predominantly positive correlation is observed in C+C collisions, with the Woods--Saxon and $\alpha$-clustered configurations remaining consistent within uncertainties. In C+Pb collisions, the two configurations exhibit a clear separation for $N_{\mathrm{part}}\gtrsim40$, with $\mathrm{NSC}\{\varepsilon_1,\varepsilon_3\}$ being smaller for the $\alpha$-clustered configuration than for the Woods--Saxon configuration. For $\mathrm{NSC}\{\varepsilon_2,\varepsilon_3\}$, negative correlations are observed in both collision systems. In C+C collisions, the two configurations are approximately consistent at low $N_{\mathrm{part}}$, up to $N_{\mathrm{part}}\sim15$, while the $\alpha$-clustered configuration becomes less negative than the Woods--Saxon configuration at larger $N_{\mathrm{part}}$. A similar configuration dependence is observed in C+Pb collisions, where the two configurations are approximately consistent at lower $N_{\mathrm{part}}$ but begin to separate around $N_{\mathrm{part}}\sim40$, with the $\alpha$-clustered configuration exhibiting a more negative $\mathrm{NSC}\{\varepsilon_2,\varepsilon_3\}$ than the Woods--Saxon configuration. These results indicate that correlations between different eccentricity harmonics can be sensitive to the initial nuclear configuration, with the effect being particularly pronounced in C+Pb collisions.


\subsection{Bulk observables with final-state particles}

All observables defined in previous section are constructed solely from the initial-state nucleon positions and thus directly reflect the geometric properties of the participant configuration. The impact of these initial-state features on the collision dynamics is subsequently assessed through final-state observables, including the yield, mean transverse momentum of identified particles, flow coefficients $v_n$, and correlations between $v_n$ and symmetric cumulants. This combined analysis allows for a systematic investigation of how initial-state geometry and clustering effects are transmitted to experimentally accessible final-state observables.

First, the bulk observables of the produced particles are examined to assess the influence of $\alpha$ clustering on final-state observables and their response to the initial-state geometry. The pseudorapidity distributions, $dN/d\eta$, for protons and pions in C+C and C+Pb collisions are shown in Fig.~\ref{fig_dndeta}. Results are presented for both the Woods--Saxon and $\alpha$-clustered configurations. The corresponding average transverse momentum, $\langle p_T \rangle$, distributions are displayed in Fig.~\ref{fig_meanptEta}. In C+C collisions, $\langle p_T \rangle$ is symmetric about mid-rapidity, as expected for a symmetric system. In asymmetric C+Pb collisions, the $\langle p_T \rangle$ distributions exhibit the expected asymmetry. When integrated over $N_{\mathrm{part}}$, the bulk observables are found to be similar for both initial-state configurations.

\begin{figure*}
\includegraphics[scale=0.65]{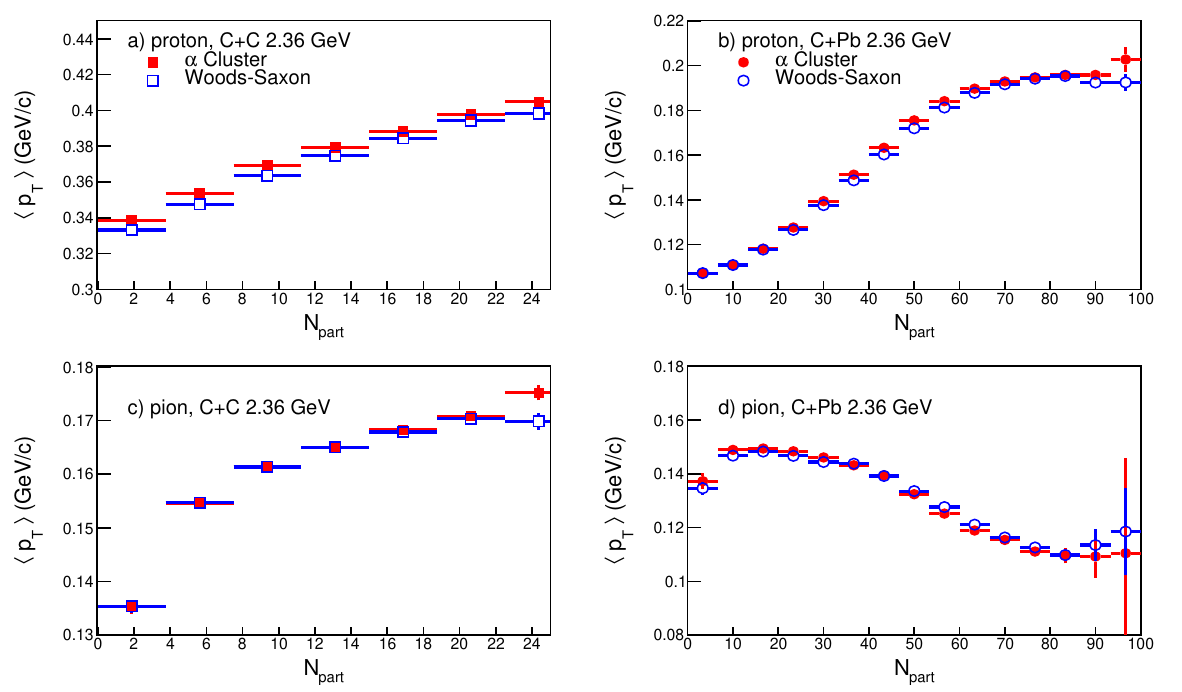}
\caption{$\langle p_T \rangle$ versus $N_{\mathrm{part}}$ for $|\eta|<2.0$ of protons and pions for Woods--Saxon and triangular $\alpha$-clustered initial configurations for C+C and C+Pb collisions at $\sqrt{s_{\mathrm{NN}}}=2.36$~GeV.
}
\label{fig_meanptNpart}
\end{figure*} 

\begin{figure*}
\includegraphics[scale=0.65]{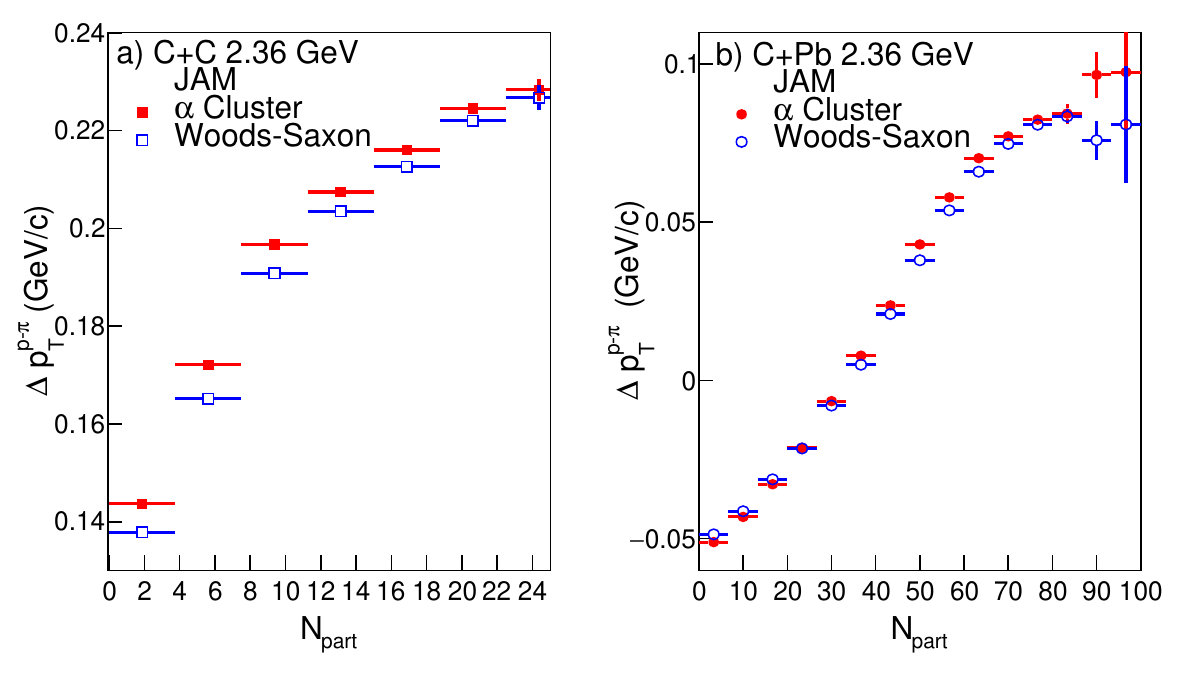}
\caption{$\Delta p_T^{(p-\pi)}$ versus $N_{\mathrm{part}}$ for $|\eta|<2.0$ for Woods--Saxon and triangular $\alpha$-clustered initial configurations for C+C and C+Pb collisions at $\sqrt{s_{\mathrm{NN}}}=2.36$~GeV.
}
\label{fig_Delta_meanptNpart}
\end{figure*} 

\subsection{Correlation in final-state particles}
\subsubsection{transverse momentum correlation}
The observed hierarchy in participant-zone compactness and transverse-size fluctuations provides a natural geometric explanation for the stronger sensitivity of mean radial observables, such as the proton $\langle p_T\rangle$, to clustering effects compared with fluctuation-driven observables.

The event-wise mean transverse momentum of identified particles is defined as~\cite{Nielsen:2023znu,Bhatta:2021qfk,Zhang:2025yyd,ATLAS:2024jvf,ALICE:2024apz}
\begin{equation}
[p_T]_\alpha
=
\frac{Q_{p_T,\alpha}}{M_\alpha}
=
\frac{1}{M_\alpha}
\sum_{i=1}^{M_\alpha} p_{T,i},
\end{equation}
where $Q_{p_T,\alpha}$ is the total transverse momentum and $M_\alpha$ is the multiplicity of particle species $\alpha$ in a given event. Here, $[\cdots]$ denotes an average over particles within a single event, while $\langle\cdots\rangle$ denotes an average over all events belonging to the same $N_{\mathrm{part}}$ interval.

The corresponding inclusive mean transverse momentum is
\begin{equation}
\langle p_T\rangle_\alpha
=
\frac{\sum_{\rm events} Q_{p_T,\alpha}}
{\sum_{\rm events} M_\alpha},
\end{equation}
where the sums run over all events in the selected $N_{\mathrm{part}}$ interval.

Transverse-momentum fluctuations are quantified by the two-particle transverse-momentum correlator
\begin{equation}
c_{2,\alpha}
=
\frac{
\sum_{\rm events}
\left(
Q_{p_T,\alpha}^{2}
-
\sum_{i=1}^{M_\alpha} p_{T,i}^{2}
\right)}
{\sum_{\rm events} M_\alpha(M_\alpha-1)}
-
\langle p_T\rangle_\alpha^{2},
\end{equation}
which quantifies the dynamical covariance of the transverse momenta of distinct particle pairs within a fixed $N_{\mathrm{part}}$ interval. For convenience, I also define the normalized fluctuation measure
\begin{equation}
k_{2,\alpha}
=
\frac{\mathrm{sign}(c_{2,\alpha})\sqrt{|c_{2,\alpha}|}}
{\langle p_T\rangle_\alpha}.
\end{equation}

To compare the mean transverse momenta of different particle species on an event-by-event basis, I also study the event-ensemble average of the difference between the proton and pion event-wise mean transverse momenta,
\begin{equation}
\Delta p_T^{(p-\pi)}
=
\left\langle
[p_T]_p
-
[p_T]_\pi
\right\rangle,
\end{equation}
where the average is taken over all events within a fixed $N_{\mathrm{part}}$ interval.


All mean transverse momentum and transverse-momentum correlation observables presented in this section are evaluated for particles within $|\eta|<2.0$. The left and right panels of Fig.~\ref{fig_meanptNpart} show $\langle p_T\rangle$ for protons and pions as a function of $N_{\mathrm{part}}$ in C+C and C+Pb collisions, respectively. In both collision systems, protons exhibit a larger $\langle p_T\rangle$ for the $\alpha$-clustered configuration than for the Woods--Saxon configuration, whereas pions show no significant difference between the two configurations. This pattern is consistent with expectations based on the hierarchy of participant-zone compactness. However, for pions, a decrease in $\langle p_T\rangle$ is observed at large $N_{\mathrm{part}}$ in C+Pb collisions. This behavior may be related to the fact that, once essentially all projectile $^{12}$C nucleons participate, the collision geometry changes primarily on the target side, while the projectile contribution remains nearly constant. This geometric saturation is accompanied by a saturation of the pion multiplicity (see Appendix~\ref{app:multiplicity}) and is consistent with the observed decrease of the pion mean transverse momentum with increasing $N_{\mathrm{part}}$.

The difference between the proton and pion mean transverse momenta, $\Delta p_T^{(p-\pi)}$, is shown in Fig.~\ref{fig_Delta_meanptNpart}. A clear distinction between the two initial-state configurations is observed in C+C collisions, while in C+Pb collisions the effect is pronounced in the mid-$N_{\mathrm{part}}$ region.

\begin{figure*}
\includegraphics[scale=0.65]{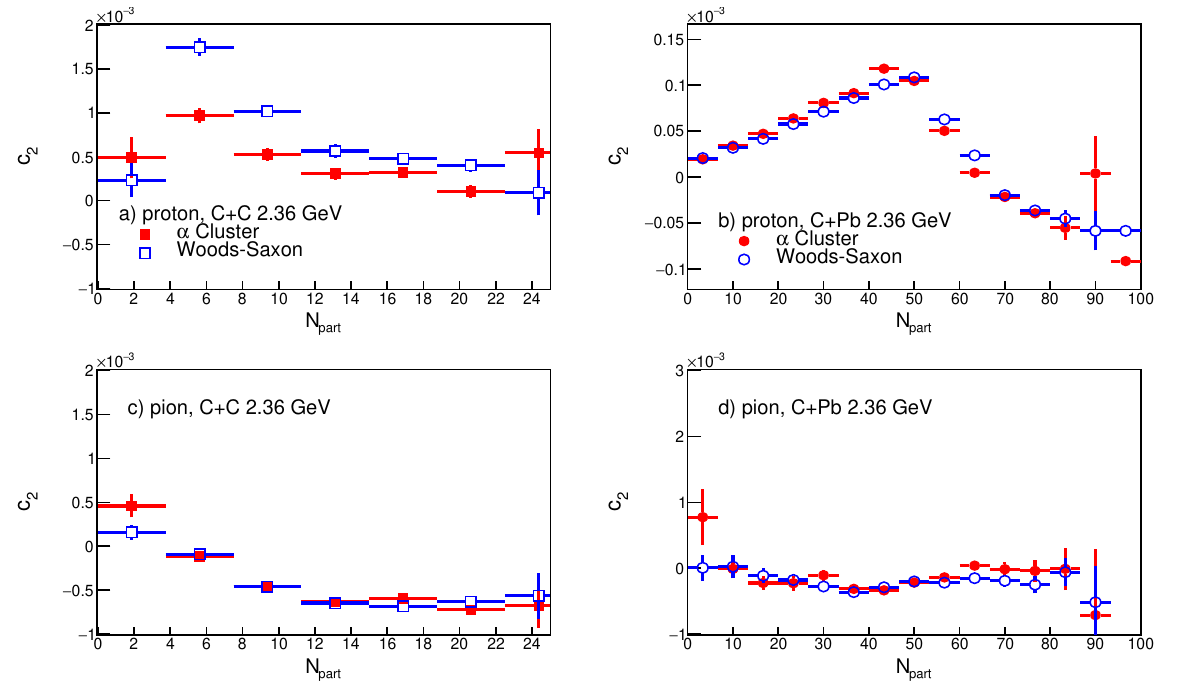}
\caption{$c_{2}$ for protons and pions as a function of $N_{\mathrm{part}}$ for Woods--Saxon and triangular $\alpha$-clustered initial configurations for C+C and C+Pb collisions at $\sqrt{s_{\mathrm{NN}}}=2.36$~GeV.
}
\label{fig_c2_proton_pion}
\end{figure*}

\begin{figure*}
\includegraphics[scale=0.65]{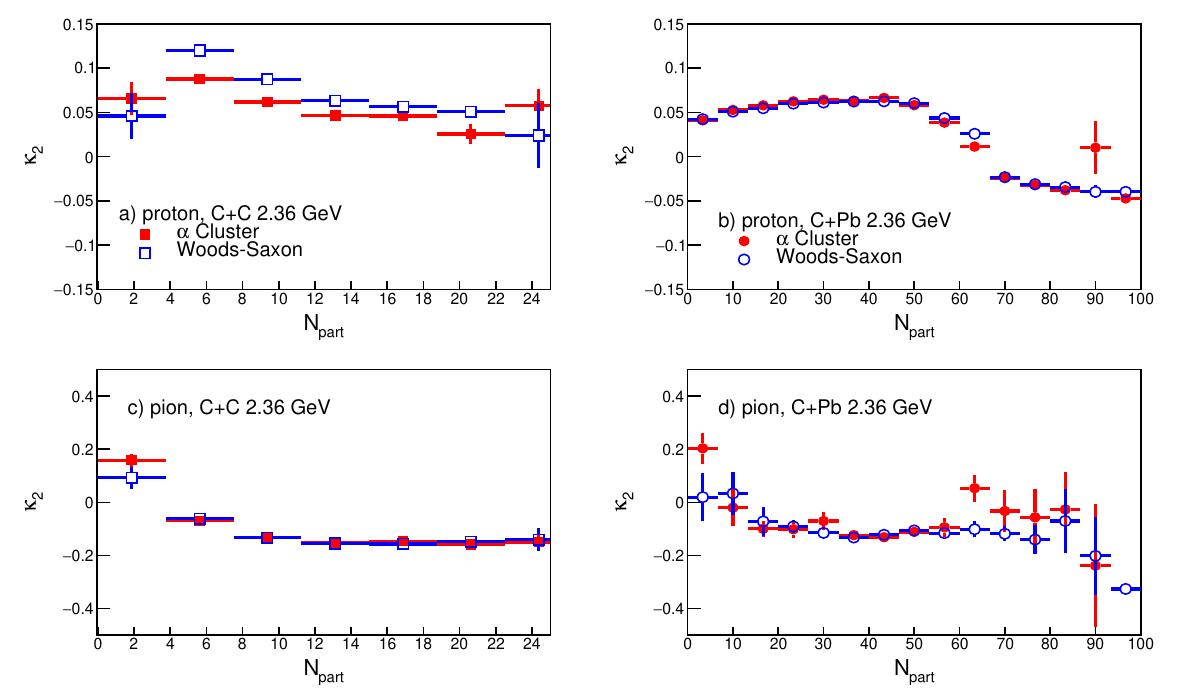}
\caption{$k_{2}$ for protons and pions as a function of $N_{\mathrm{part}}$ for Woods--Saxon and triangular $\alpha$-clustered initial configurations for C+C and C+Pb collisions at $\sqrt{s_{\mathrm{NN}}}=2.36$~GeV.
}
\label{fig_k2_proton_pion}
\end{figure*}

The proton and pion transverse-momentum fluctuation observables, $c_{2}$ and $k_{2}$, for C+C and C+Pb collisions are presented in Figs.~\ref{fig_c2_proton_pion} and~\ref{fig_k2_proton_pion}, respectively. In C+C collisions, the proton correlator $c_{2,p}$ is systematically smaller for the $\alpha$-clustered configuration than for the Woods--Saxon configuration, consistent with the ordering observed for the transverse-size fluctuations of the participant zone. No such clear ordering is observed in C+Pb collisions. In particular, for C+Pb collisions, $c_{2,p}$ increases with $N_{\mathrm{part}}$, reaches a maximum around $N_{\mathrm{part}}\sim50$, and then decreases, becoming negative for $N_{\mathrm{part}}\gtrsim70$. The normalized correlator $k_{2,p}$ exhibits a similar dependence on $N_{\mathrm{part}}$.

To investigate the origin of this sign change, I examined the participant composition, as well as the average proton multiplicity and its variance, as functions of $N_{\mathrm{part}}$ (see Appendix~\ref{app:multiplicity}). In C+Pb collisions, the mean number of C participants increases with $N_{\mathrm{part}}$ and approaches the full carbon-nucleus size, $\langle N_{\mathrm{part}}^{C}\rangle\simeq12$, around $N_{\mathrm{part}}\sim 40$, while $\langle N_{\mathrm{part}}^{Pb}\rangle$ continues to increase with $N_{\mathrm{part}}$. Thus, $N_{\mathrm{part}}\sim 40$ marks a transition to a regime in which the C nucleus is essentially fully participating and further increases in $N_{\mathrm{part}}$ are predominantly due to additional Pb participants. In the same region, the proton multiplicity variance, $\mathrm{Var}(M_p)$, reaches a maximum and decreases thereafter, while the mean proton multiplicity continues to increase and gradually saturates around $N_{\mathrm{part}}\sim70$. The coincidence of the change in participant composition with the maximum in $\mathrm{Var}(M_p)$ suggests that the observed structure in the transverse-momentum correlator is related to the finite and asymmetric geometry of the C+Pb system. As the system evolves toward larger $N_{\mathrm{part}}$, the reduction of multiplicity fluctuations can suppress the positive contribution to the transverse-momentum correlator. Consequently, negative correlations associated with local conservation laws in the proton-dominated final state may become increasingly important for $N_{\mathrm{part}}\gtrsim70$. This behavior is not observed in the smaller, symmetric C+C system.

In contrast, the pion correlators, $c_{2,\pi}$ and $k_{2,\pi}$, exhibit only a weak dependence on $N_{\mathrm{part}}$ and show no significant difference between the Woods--Saxon and $\alpha$-clustered initial configurations.

\subsubsection{flow correlation}

To characterize the magnitude and event-by-event fluctuations of the anisotropic flow, I employ the standard $Q$-vector cumulant formalism~\cite{Bilandzic:2010jr,Borghini:2001vi,ALICE:2016kpq}. For each event, the flow vector of harmonic order $n$ is defined as
\begin{equation}
Q_n
=
\sum_{i=1}^{M}
e^{in\phi_i},
\end{equation}
where $M$ is the multiplicity of the selected charged particles and $\phi_i$ is the azimuthal angle of particle $i$. The corresponding event-wise two-particle azimuthal correlator is
\begin{equation}
[2]_n
=
\frac{|Q_n|^2-M}
{M(M-1)},
\end{equation}
where $[\cdots]$ denotes an average over particle pairs within a single event.

The two-particle flow cumulant is obtained from the weighted average over events,
\begin{equation}
v_n\{2\}
=
\sqrt{\langle [2]_n\rangle},
\end{equation}
where $\langle\cdots\rangle$ denotes an average over events at fixed $N_{\mathrm{part}}$, weighted by the number of particle pairs, $M(M-1)$. I note that the interpretation of $v_1\{2\}$ requires particular caution. The initial dipole eccentricity $\varepsilon_1$ is associated with the rapidity-even component of directed flow, $v_{1,\mathrm{even}}$. However, the inclusive two-particle correlator used to calculate $v_1\{2\}$ does not isolate this component and can also receive contributions from global momentum conservation (GMC), which may be comparable in magnitude to the genuine rapidity-even flow signal~\cite{Magdy:2023fsp,Magdy:2026tfh}. Since the present analysis does not include a dedicated treatment of GMC or a separation of the rapidity-even and rapidity-odd components, the extracted $v_1\{2\}$ cannot be regarded as a reliable measurement of $v_{1,\mathrm{even}}$. A dedicated analysis to isolate the genuine rapidity-even component, account for GMC, and investigate its relationship with the initial-state $\varepsilon_1$ is beyond the scope of the present work and is left for future study.

To quantify event-by-event fluctuations of the two-particle correlator, I evaluate its variance,
\begin{equation}
\mathrm{Var}([2]_n)
=
\langle [2]_n^2\rangle
-
\langle [2]_n\rangle^2,
\end{equation}
where the averages are taken over all events within a fixed $N_{\mathrm{part}}$ interval.

The correlations between different flow harmonics are quantified by the symmetric cumulants,
\begin{equation}
\mathrm{SC}\{v_m,v_n\}
=
\langle [4]_{m,n}\rangle
-
\langle [2]_m\rangle
\langle [2]_n\rangle,
\end{equation}
where $[4]_{m,n}$ denotes the event-wise mixed-harmonic four-particle correlator evaluated using the standard $Q$-vector cumulant formalism~\cite{Bilandzic:2010jr,ALICE:2016kpq}.

\begin{figure*}
\includegraphics[scale=0.65]{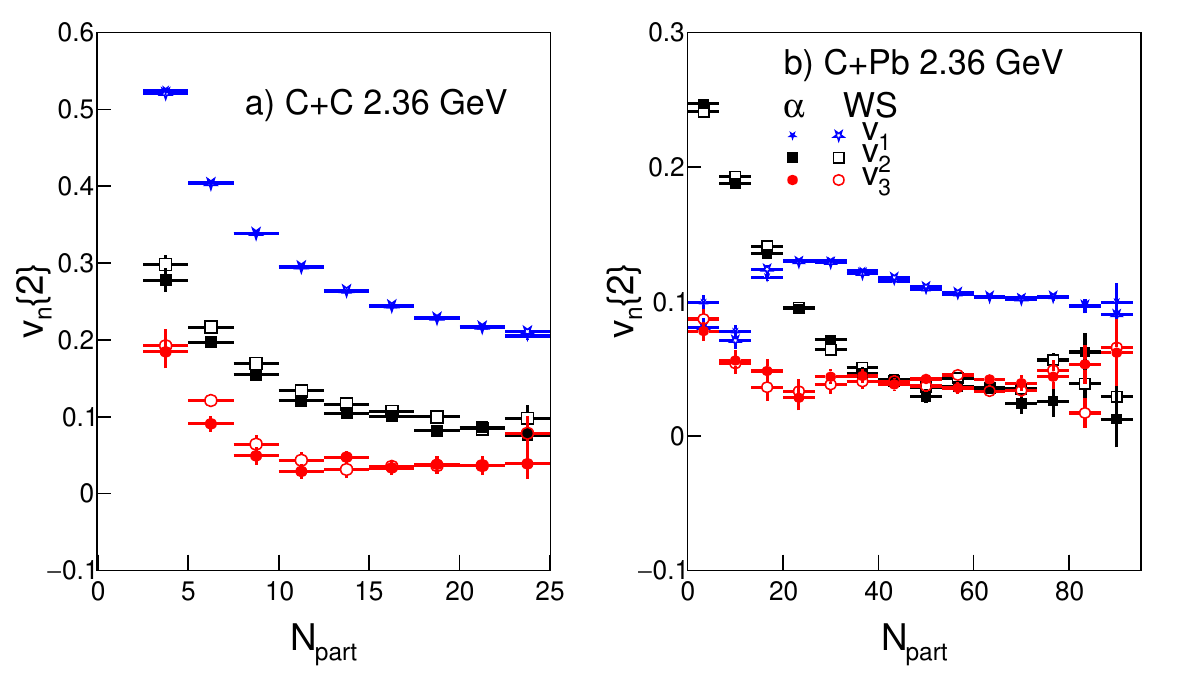}
\caption{$v_n\{2\}$ as a function of
$N_{\mathrm{part}}$ for Woods--Saxon and triangular $\alpha$-clustered initial
configurations for C+C and C+Pb collisions at $\sqrt{s_{\mathrm{NN}}}=2.36$~GeV.
}
\label{fig_vn_cumu_v123}
\end{figure*} 

\begin{figure*}
\includegraphics[scale=0.65]{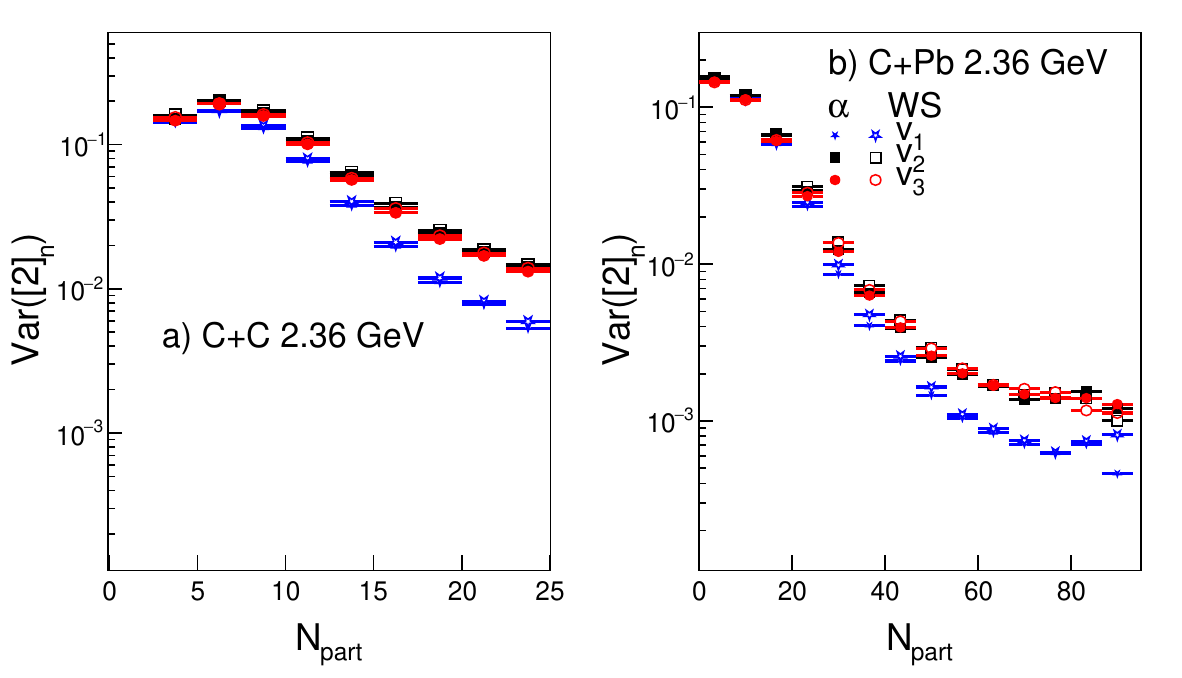}
\caption{$\mathrm{Var}([2]_n)$ as a function of
$N_{\mathrm{part}}$ for Woods--Saxon and triangular $\alpha$-clustered initial
configurations for C+C and C+Pb collisions at $\sqrt{s_{\mathrm{NN}}}=2.36$~GeV.
}
\label{fig_vn_cumuVar_v123}
\end{figure*} 

All anisotropic-flow and flow-correlation observables presented in this section are evaluated using all charged particles with $|\eta|<2.0$ and $0.2<p_T<2.0$~GeV/$c$. Figures~\ref{fig_vn_cumu_v123} present $v_n\{2\}$ as a function of $N_{\mathrm{part}}$ for the Woods--Saxon and triangular $\alpha$-clustered initial configurations in C+C and C+Pb collisions. In C+C collisions, all three flow harmonics decrease with increasing $N_{\mathrm{part}}$, following the ordering $v_1\{2\} > v_2\{2\} > v_3\{2\}$. The interpretation of $v_1\{2\}$, however, requires caution because the inclusive two-particle correlator can receive contributions from global momentum conservation (GMC) and does not isolate the genuine rapidity-even directed-flow component. The ordering $v_2\{2\} > v_3\{2\}$ is consistent with that of the corresponding initial eccentricities, $\varepsilon_2 > \varepsilon_3$, although the final-state flow coefficients also depend on the dynamical response of the collision system. Both the initial eccentricities and the flow harmonics exhibit only a weak dependence on the choice of the initial-state configuration. In contrast, C+Pb collisions exhibit a non-trivial dependence on $N_{\mathrm{part}}$. In these collisions, $v_1\{2\}$ and $v_3\{2\}$ remain approximately independent of $N_{\mathrm{part}}$, whereas $v_2\{2\}$ decreases with increasing $N_{\mathrm{part}}$ at lower values and saturates for $N_{\mathrm{part}}\gtrsim40$. At $\sqrt{s_{\mathrm{NN}}}=2.36$~GeV, baryon stopping, finite net-baryon density, mean-field interactions, and hadronic rescattering can influence the relationship between the initial collision geometry and the final-state anisotropic flow. The weak sensitivity of the measured flow coefficients to the initial-state configuration suggests that the differences between the Woods--Saxon and triangular $\alpha$-clustered geometries are not strongly reflected in the final-state observables considered here.

The variances of the event-wise two-particle correlators, $\mathrm{Var}([2]_1)$, $\mathrm{Var}([2]_2)$, and $\mathrm{Var}([2]_3)$, also decrease approximately with increasing $N_{\mathrm{part}}$ in both collision systems. Similar to the corresponding average flow coefficients, no appreciable differences are observed between the $\alpha$-clustered and Woods--Saxon configurations for any of the flow harmonics.

\begin{figure*}
\includegraphics[scale=0.65]{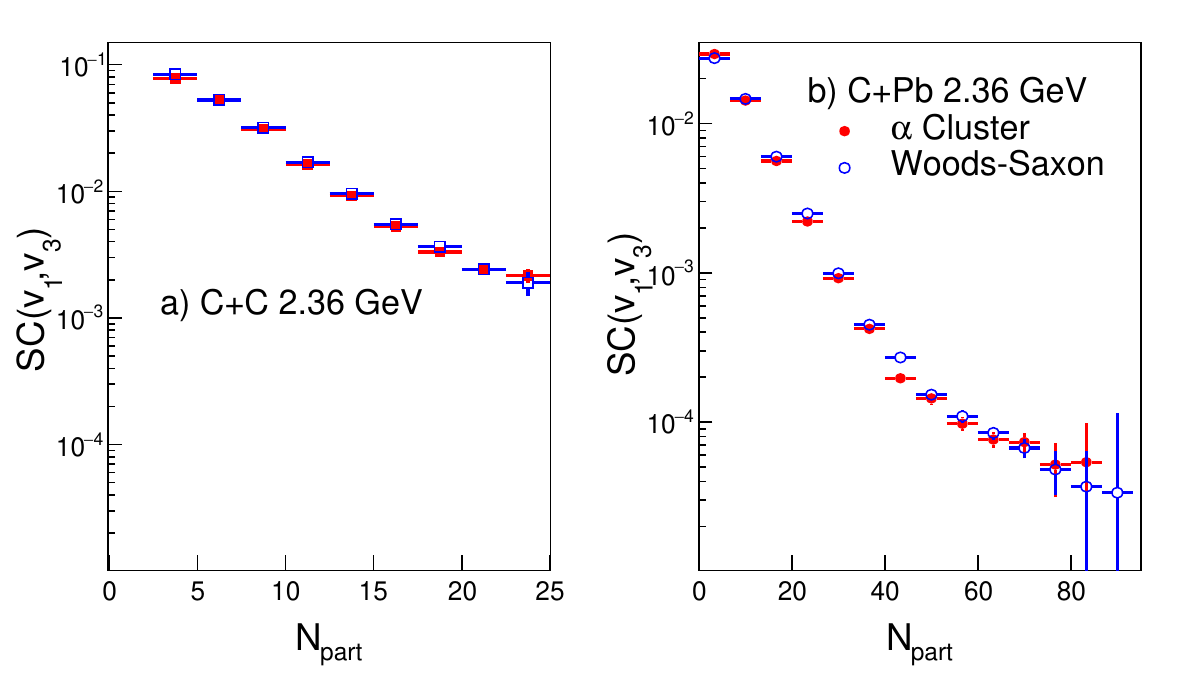}
\caption{$\mathrm{SC}\{v_1,v_3\}$ as a function of
$N_{\mathrm{part}}$ for Woods--Saxon and triangular $\alpha$-clustered initial
configurations for C+C and C+Pb collisions at $\sqrt{s_{\mathrm{NN}}}=2.36$~GeV.
}
\label{fig_SC13_vn}
\end{figure*}

\begin{figure*}
\includegraphics[scale=0.65]{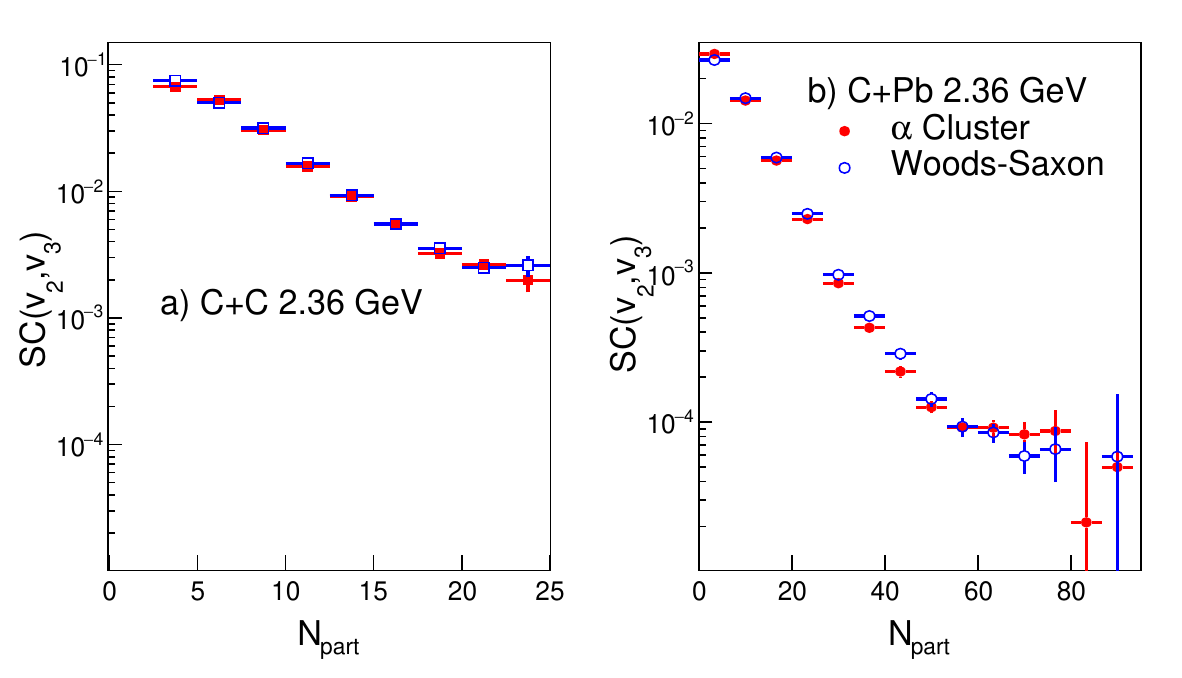}
\caption{$\mathrm{SC}\{v_2,v_3\}$ as a function of
$N_{\mathrm{part}}$ for Woods--Saxon and triangular $\alpha$-clustered initial
configurations for C+C and C+Pb collisions at $\sqrt{s_{\mathrm{NN}}}=2.36$~GeV.
}
\label{fig_SC23_vn}
\end{figure*}

Finally, the symmetric cumulants of the flow harmonics, $\mathrm{SC}\{v_1,v_3\}$ and $\mathrm{SC}\{v_2,v_3\}$, are presented in Figs.~\ref{fig_SC13_vn} and~\ref{fig_SC23_vn}, respectively. Both symmetric cumulants remain positive over the entire $N_{\mathrm{part}}$ range in C+Pb and C+C collisions. In both systems, they exhibit a similar decreasing trend with increasing $N_{\mathrm{part}}$ and have comparable magnitudes. No appreciable differences are observed between the Woods--Saxon and $\alpha$-clustered initial configurations for either correlation. In contrast to the corresponding initial-state eccentricity correlations, the flow symmetric cumulants show little sensitivity to $\alpha$ clustering. This indicates that the sensitivity to the initial nuclear configuration observed in the eccentricity correlations is not comparably evident in these final-state flow observables at $\sqrt{s_{\mathrm{NN}}}=2.36$~GeV. I also evaluated the corresponding normalized symmetric cumulants, $\mathrm{NSC}\{v_m,v_n\}=\mathrm{SC}\{v_m,v_n\}/(\langle [2]_m\rangle\langle [2]_n\rangle)$. However, at this collision energy, the two-particle flow correlators $\langle [2]_m\rangle$ and $\langle [2]_n\rangle$ become very small, making the normalization statistically unstable and resulting in large fluctuations in the extracted $\mathrm{NSC}\{v_m,v_n\}$. Therefore, the normalized flow cumulants are not shown or used as additional final-state observables.

\section{Conclusion}

In this work, I have investigated the sensitivity of low-energy nuclear collisions to intrinsic nuclear structure by studying C+C and C+Pb collisions at $\sqrt{s_{\mathrm{NN}}}=2.36$~GeV within the JAM transport model. By comparing Woods--Saxon and triangular $\alpha$-clustered configurations for the $^{12}\mathrm{C}$ nucleus, I systematically explored how initial-state geometric features propagate to final-state observables in a regime dominated by baryon stopping, mean-field interactions, and hadronic rescattering.

I find that $\alpha$ clustering leads to a more compact participant configuration than the Woods--Saxon case, while the average eccentricities and their fluctuations exhibit only a weak sensitivity to the initial-state configuration. The root-mean-square radii of the two configurations differ by less than a few percent, indicating that the observed effects are not driven by trivial changes in nuclear size, but rather by differences in the microscopic spatial organization of nucleons. At this beam energy, radial observables remain directly sensitive to such geometric compactness. In particular, the proton mean transverse momentum is enhanced for the $\alpha$-clustered configuration, whereas pion observables show little response. The proton transverse-momentum correlators, $c_{2}$ and $k_{2}$, are also sensitive to the initial-state geometry and follow the hierarchy of the participant-zone transverse-size fluctuations, while the corresponding pion observables remain largely unaffected.

The collective anisotropic response of the system was investigated through two-particle flow cumulants and symmetric cumulants. Although the average flow harmonics exhibit systematic centrality dependence, no appreciable differences are observed between the Woods--Saxon and $\alpha$-clustered configurations. Likewise, the event-by-event fluctuations of the flow harmonics and the final-state symmetric cumulants show little sensitivity to the initial-state configuration. In contrast, the correlations between different harmonic components of the initial geometry, quantified by the normalized symmetric cumulants $\mathrm{NSC}\{\varepsilon_m,\varepsilon_n\}$, exhibit a clear dependence on $\alpha$ clustering, particularly for $\mathrm{NSC}\{\varepsilon_2,\varepsilon_3\}$ in C+Pb collisions. These results indicate that, at $\sqrt{s_{\mathrm{NN}}}=2.36$~GeV, the sensitivity to $\alpha$ clustering observed in the initial-state eccentricity correlations is not comparably evident in the final-state anisotropic-flow observables.

Overall, our results show that low-energy nuclear collisions retain measurable sensitivity to $\alpha$ clustering primarily through radial observables, while signatures in anisotropic-flow observables are considerably reduced by the subsequent dynamical evolution. I note, however, that the $\alpha$-clustered configuration employed in this work represents a simplified geometric implementation, and that more realistic descriptions incorporating state-of-the-art nuclear structure calculations from low-energy nuclear physics will be essential for quantitative constraints on clustering effects. These findings provide a baseline for future studies of nuclear-structure effects in low-energy heavy-ion collisions and motivate corresponding experimental measurements at CSR and HIAF.

\section*{\label{sec:acknowledgments}Acknowledgments}
SS thanks the colleagues at the Institute of Modern Physics for valuable discussions. SS would also like to thank the anonymous referee for providing excellent suggestions that led to improvements in the analysis. SS also thanks Dr.\ Amaresh Jaiswal and Dr.\ Sandeep Chatterjee for useful discussions during the WHEPP 2025 conference. SS acknowledges support from the Chinese
Academy of Sciences during this work.

\appendix
\section{Participant Composition and Multiplicity Fluctuations}
\label{app:multiplicity}

\begin{figure*}
\includegraphics[scale=0.3]{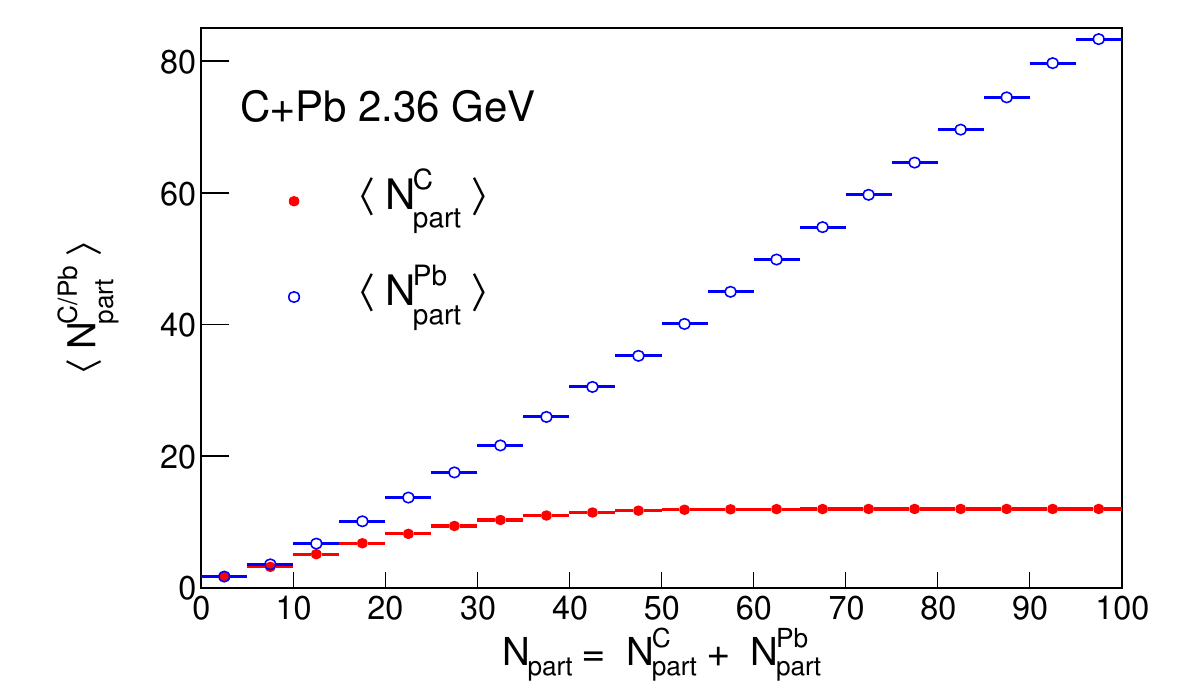}
\caption{Average numbers of participants from the C and Pb nuclei,
$\langle N_{\mathrm{part}}^{C}\rangle$ and $\langle N_{\mathrm{part}}^{Pb}\rangle$,
as functions of the total number of participants, $N_{\mathrm{part}}$,
in C+Pb collisions at $\sqrt{s_{\mathrm{NN}}}=2.36$~GeV for the
Woods--Saxon and triangular $\alpha$-clustered initial configurations.}
\label{fig_appen_participant}
\end{figure*}

Figure~\ref{fig_appen_participant} shows the average numbers of participants from the C and Pb nuclei as functions of $N_{\mathrm{part}}$ in C+Pb collisions at $\sqrt{s_{\mathrm{NN}}}=2.36$~GeV. The total number of participants is given by
\begin{equation}
N_{\mathrm{part}}=N_{\mathrm{part}}^{C}+N_{\mathrm{part}}^{Pb}.
\end{equation}
The mean number of C participants increases with $N_{\mathrm{part}}$ and approaches the full carbon-nucleus size, $\langle N_{\mathrm{part}}^{C}\rangle\simeq12$, around $N_{\mathrm{part}}\sim40$, while $\langle N_{\mathrm{part}}^{Pb}\rangle$ continues to increase with $N_{\mathrm{part}}$. This indicates a transition around $N_{\mathrm{part}}\sim40$ from increasing participation of the C nucleus to a regime in which the C nucleus is essentially fully participating and further increases in $N_{\mathrm{part}}$ are predominantly due to additional Pb participants.

\begin{figure*}
\includegraphics[scale=0.5]{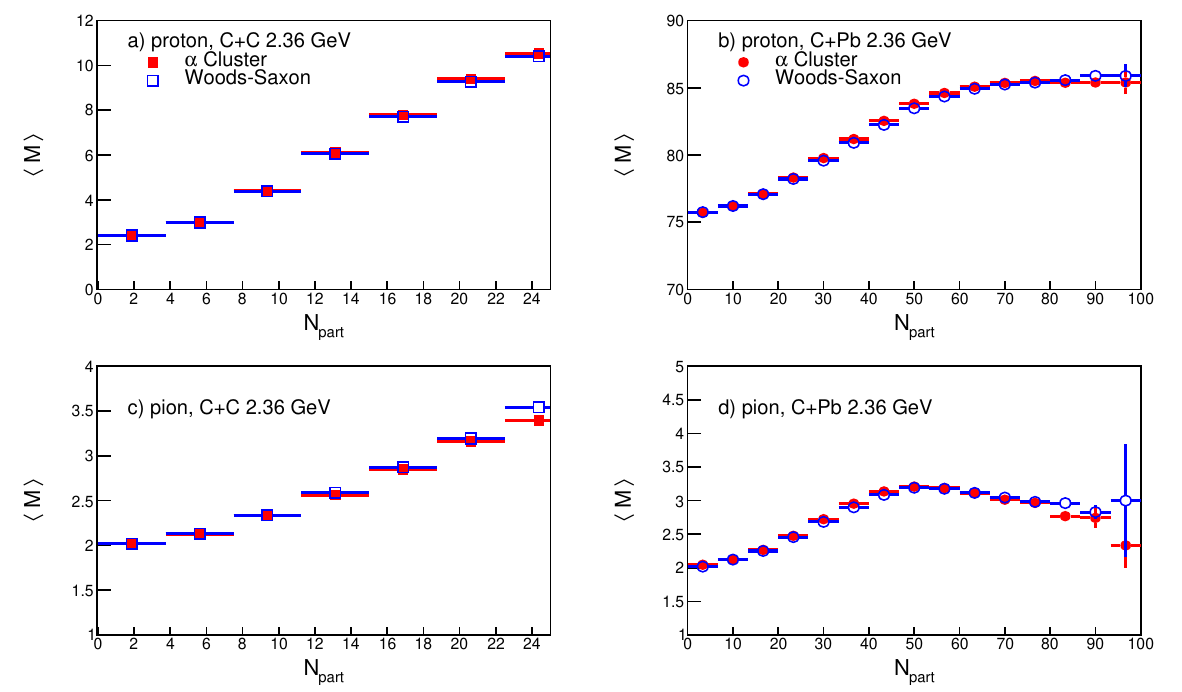}
\caption{Average accepted final-state proton and pion multiplicities,
$\langle M_p\rangle$ and $\langle M_{\pi}\rangle$, as functions of
$N_{\mathrm{part}}$ for Woods--Saxon and triangular $\alpha$-clustered
initial configurations in C+C and C+Pb collisions at
$\sqrt{s_{\mathrm{NN}}}=2.36$~GeV.}
\label{fig_appen_multNpart}
\end{figure*}

\begin{figure*}
\includegraphics[scale=0.5]{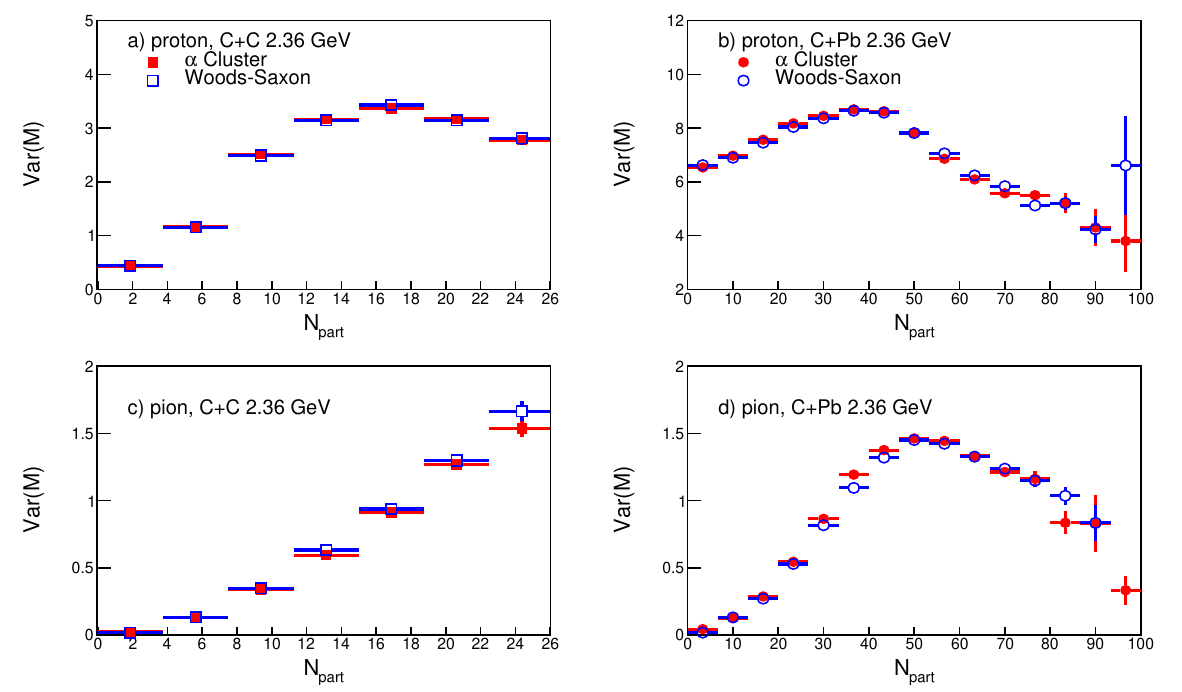}
\caption{Variances of the accepted final-state proton and pion
multiplicities, $\mathrm{Var}(M_p)$ and $\mathrm{Var}(M_{\pi})$, as
functions of $N_{\mathrm{part}}$ for Woods--Saxon and triangular
$\alpha$-clustered initial configurations in C+C and C+Pb collisions
at $\sqrt{s_{\mathrm{NN}}}=2.36$~GeV.}
\label{fig_appen_multVarNpart}
\end{figure*}

Figures~\ref{fig_appen_multNpart} and~\ref{fig_appen_multVarNpart} present the average accepted final-state proton and pion multiplicities, $\langle M_p\rangle$ and $\langle M_{\pi}\rangle$, and their variances, $\mathrm{Var}(M_p)$ and $\mathrm{Var}(M_{\pi})$, as functions of $N_{\mathrm{part}}$ for C+C and C+Pb collisions. Results are shown for both the Woods--Saxon and triangular $\alpha$-clustered initial configurations. In C+Pb collisions, the proton multiplicity variance reaches a maximum around $N_{\mathrm{part}}\sim40$ and decreases thereafter, while the mean proton multiplicity continues to increase and gradually saturates around $N_{\mathrm{part}}\sim70$. These features provide additional information on the multiplicity fluctuations associated with the finite and asymmetric geometry of the C+Pb system.

\bibliography{reference}
\end{document}